\documentclass{emulateapj}
\usepackage{epsfig}
\usepackage{epsf}
\newcommand{\beq}{\begin{equation}}
\newcommand{\eeq}{\end{equation}}

\def\comment#1{\par\noindent\llap{$\Rightarrow$\enskip}{\bf #1}\par}


\def \logTd6 {\hbox{log$( T/6 \kev)$} }

\def\myputfigure#1#2#3#4#5%
{\vskip#5pt\makebox[0pt]{\hskip#2in
\includegraphics[width=#3\textwidth]{#1}}\vskip#4pt\hfill}

\def \ie        {\hbox{\it i.e.}}
\def \eg        {\hbox{\it e.g.}}

\def \etal      {et al.\ }

\def \kev       {{\rm\ keV}}

\def \hMpc      {h^{-1}{\rm\ Mpc}}

\bibpunct{(}{)}{;}{a}{,}{,}
\input epsf



\def\hatn{{\bf \hat n}}
\def\hatk{{\bf \hat k}}
\def\hatq{{\bf \hat q}}

\def\vecq{{\bf q}}

\def\veck{{\bf k}}
\def\vecv{{\bf v}}
\def\vecp{{\bf p}}
\def\vecn{{\bf n}}

\def\hatk{{\bf \hat k}}

\def\hatq{{\bf \hat q}}
\def\neff{n_{\rm eff}}
\def\Feff{F^{\rm eff}_{2}}
\def\norm{\left|\veck + \vecq\right|}

\long\def\comment#1{}

\def\etal{{\it et al.~}}

\def\bfell{{\mbox{\boldmath $\ell$}}}

\def\W2{{\cal W}}

\def\be{\begin{equation}}
\def\ee{\end{equation}}
\def\bea{\begin{eqnarray}}
\def\eea{\end{eqnarray}}

\def\l{{\ell}}

\def\cmm2{{\,\rm cm^{-2}}}
\def\cm2{{\,{\rm cm}^2}}
\def\cmm3{{\,{\rm cm}^{-3}}}
\def\gcmm3{{\,{\rm g\,cm^{-3}}}}

\def\fun#1#2{\lower3.6pt\vbox{\baselineskip0pt\lineskip.9pt
  \ialign{$\mathsurround=0pt#1\hfil##\hfil$\crcr#2\crcr\sim\crcr}}}

\def \eg {{\it e.g. }}
\def \ie {{\it i.e. }}

\lefthead{ Dor\'e, Hennawi, \& Spergel}
\righthead{Beyond the Damping Tail}

\begin{document}
\bibliographystyle{apj}

\title{Beyond the Damping Tail:\\ Cross-Correlating the Kinetic Sunyaev-Zel'dovich Effect with Cosmic Shear}
\author{Olivier Dor\'e, Joseph F. Hennawi, \& David N. Spergel}
\affil{Department of Astrophysical Sciences, Peyton Hall, Ivy Lane, Princeton University, NJ-08544, USA\\
email: olivier@astro.princeton.edu, jhennawi@astro.princeton.edu, dns@astro.princeton.edu}

\begin{abstract}
Secondary anisotropies of the CMB have the potential to reveal
intricate details about the history of our universe between the 
present and recombination epochs. However, because the CMB we observe 
is the projected sum of a multitude of effects, the interpretation of
small scale anisotropies by future high resolution experiments will be 
marred by uncertainty and speculation without the handles provided by other
observations. The recent controversy over the \emph{excess} small scale
anisotropy detected by CBI and the BIMA array is a foretaste of 
potential challenges that will be faced when interpreting future experiments. In 
this paper we show that cross correlating the CMB with an overlapping 
weak lensing survey will isolate the elusive kinetic Sunyaev-Zel'dovich Effect 
from secondary anisotropies generated at higher redshifts.  We show that
if upcoming high angular resolution CMB experiments, like PLANCK/ACT/SPT, 
cover the same area of sky as current and future weak lensing surveys, 
like CFTHLS/SNAP/LSST, the cross correlation of cosmic shear with the
kSZ effect will be detected with high signal to noise ratio,
increasing the potential science accessible to both sets of surveys. 
For example, if ACT and a CFHTLS like survey were to overlap this 
cross-correlation would be detected with a total signal to noise ratio
greater than 220, reaching 1.8 per individual multipole around $\l
\sim 5000$.  Furthermore, this cross-correlation probes the three point coupling 
between the underlying dark matter and the \emph{momentum} of the 
ionized baryons in the densest regions of the universe at intermediate
redshifts. Similar to the tSZ power spectrum, its strength is
extremely sensitive to the power spectrum normalization parameter,  
$\sigma_8$, scaling roughly as $\sigma_8^7$. It provides an effective 
mechanism to isolate any component of anisotropy due to patchy reionization 
and rule out primordial small scale anisotropy.
\end{abstract}


\section{Introduction}

The Wilkinson Microwave Anisotropy Probe (WMAP)~\footnote{http://map.gsfc.nasa.gov/} 
has ushered in an era of unprecedented accuracy for measurements of the cosmic microwave background (CMB)
anisotropy.  Future high angular resolution experiments like 
PLANCK\footnote{http://astro.estec.esa.nl/SA-general/Projects/Planck}, the 
Atacama Cosmology Telescope (ACT)\footnote{{\texttt
    http://www.hep.upenn.edu/$\sim$angelica/act/act.html}}, and the South
Pole Telescope (SPT)\footnote{{\texttt http://astro.uchicago.edu/spt}} will
measure arc-minute scale ($\ell \gtrsim 1000$) anisotropies at the
$\mu$K level. The primary anisotropies generated at the epoch of
recombination and measured by WMAP \citep{Sp03} involve 
calculations in linear perturbation theory and are on sound theoretical footing 
\citep[see however][and references therein for a discussion of alternative models]{Be03}. However, 
secondary anisotropies, caused by highly nonlinear structures and involving  
complicated dissipative baryonic physics, are still a subject of theoretical 
speculation 

In considering secondary anisotropies, a practical distinction must be made 
between anisotropies that have a CMB-like thermal spectrum, and the
anisotropies with a non thermal frequency dependence, most notably the
thermal Sunyaev-Zel'dovich effect (tSZ) \citep{sunyaev80}. Although
challenging and imperfect, it should be possible to isolate
these thermal contributions from a temperature map using the 
specific frequency and spatial dependence of the latter 
\citep{BoGi99,TeEf96}. After these non-thermal components have been removed,  
the dominant sources of secondary anisotropies are expected to be 
--- from degree to arc-minute scales --- the Rees-Sciama effect 
(RS)\citep{ReSc68}, the weak gravitational lensing of the CMB itself, 
the kinetic Sunyaev-Zel'dovich (kSZ) effect and possibly patchy
reionization (Sunyaev \& Zel'dovich 1980 and Santos \etal 2003 for
recent references). Given a ``frequency cleaned'' temperature map
containing the sum of these secondary effects, it would be desirable
to study each component individually in order to evaluate  their
consistency with theoretical predictions. If the CMB  temperature were
studied in conjunction with another low redshift tracer of the
cosmological density field,  the presence or absence of correlations
between the two might provide a mechanism to isolate the various
components. 

The recent detection of \emph{excess} small scale anisotropy is a nice 
illustration of the preceding discussion and provides a foretaste of the 
potential challenges that will be faced when interpreting the secondary 
anisotropies detected by future experiments.  
At angular scales of a several arc-minutes at multipoles beyond 
the damping tail $(\ell \gtrsim 2000)$, 
the CBI experiment  \citep{Mason03} and the BIMA array \citep{Da02} 
have both detected temperature anisotropy at a level of $\sim 500 \mu$K$^2$ 
with significance of $\gtrsim 3\sigma$.  Although a natural interpretation of 
this excess power is tSZ, the strength of the signal requires that the 
power spectrum normalization parameter take on a value  
$\sigma_8 \approx 1$ in the upper range allowed by current CMB and Large 
Scale Structure data \citep{Bond03,KoSe02}. Furthermore, the required 
value is likely even higher because the measured 
anisotropy, if thermal SZ, was likely diluted by radio point source 
subtraction \citep{Ho02}.
Both the CBI and BIMA experiments observe at low frequencies 
$\sim 30$ GHz well into the Rayleigh-Jeans region of the 
thermal SZ frequency spectrum, so at present the non-thermal frequency 
signature of the tSZ effect cannot be exploited
to determine the nature of this excess. This has
led to considerable speculation about other sources of arc-minute
scale anisotropy possibly arising from the epoch of reionization or the early 
universe. 
While the kSZ effect and patchy reionization are not expected to produce 
small scale anisotropies at the level detected, 
it has been suggested that the excess power is due to SZ fluctuations 
from high-z $\gtrsim 10$ star formation \citep{Oh03}, 
primordial voids created by bubble nucleation during the inflationary 
epoch \citep{Gr03}, or broken scale invariance in the primordial power 
spectrum produced by inflation \citep{Co02}. Cross correlating 
the CMB temperature maps measured by CBI and BIMA with a low redshift 
tracer of the density field would provide a mechanism to determine whether 
the small scale excess was generated in the local universe. The detection 
of a correlation with galaxies would favor tSZ or radio source
contamination rather than the more speculative alternatives.

Cosmologists will likely face similar uncertainties in the future for 
experiments that will have the frequency coverage required to produce 
frequency cleaned temperature maps that effectively remove the tSZ. 
Cross correlation with a local density tracer could provide the most effective
means to determine the source of the anisotropy. For example, the patchy 
reionization signal or some exotic form of small scale 
anisotropy generated by inflation will not correlate with low redshift 
density tracers; whereas, the kinetic SZ effect should correlate strongly, 
which brings us to the subject of this work. Weak gravitational lensing, 
or the coherent distortion of images of faint background galaxies by the 
foreground matter distribution \citep[see][for a recent review]{vWMe03}, is a 
tracer of the local density field, with the added advantage that it 
probes the dark matter directly, foregoing the complications caused by 
issues of bias in galaxy surveys. In this paper we compute the correlation 
between weak gravitational lensing and the kinetic Sunyaev-Zel'dovich effect
and evaluate the prospects of future experiments to measure it. 

Although the cross-correlations between secondary anisotropies and weak 
gravitational lensing of the CMB itself, has been considered
previously (Spergel \& Goldberg 1999, Goldberg \& Spergel 1999, Cooray
\& Hu 2000, Takada \& Sugyama 2002, Verde \& Spergel 2002), the potential
correlation signals between the CMB and galaxy weak lensing, 
or ``cosmic shear'' has been somewhat neglected.  In particular, 
we focus on previously unexplored small angular scales beyond the 
``damping tail,'' which will be probed by the aforementioned future 
experiments.  

\begin{figure*}[t]
  \centerline{\epsfig{file =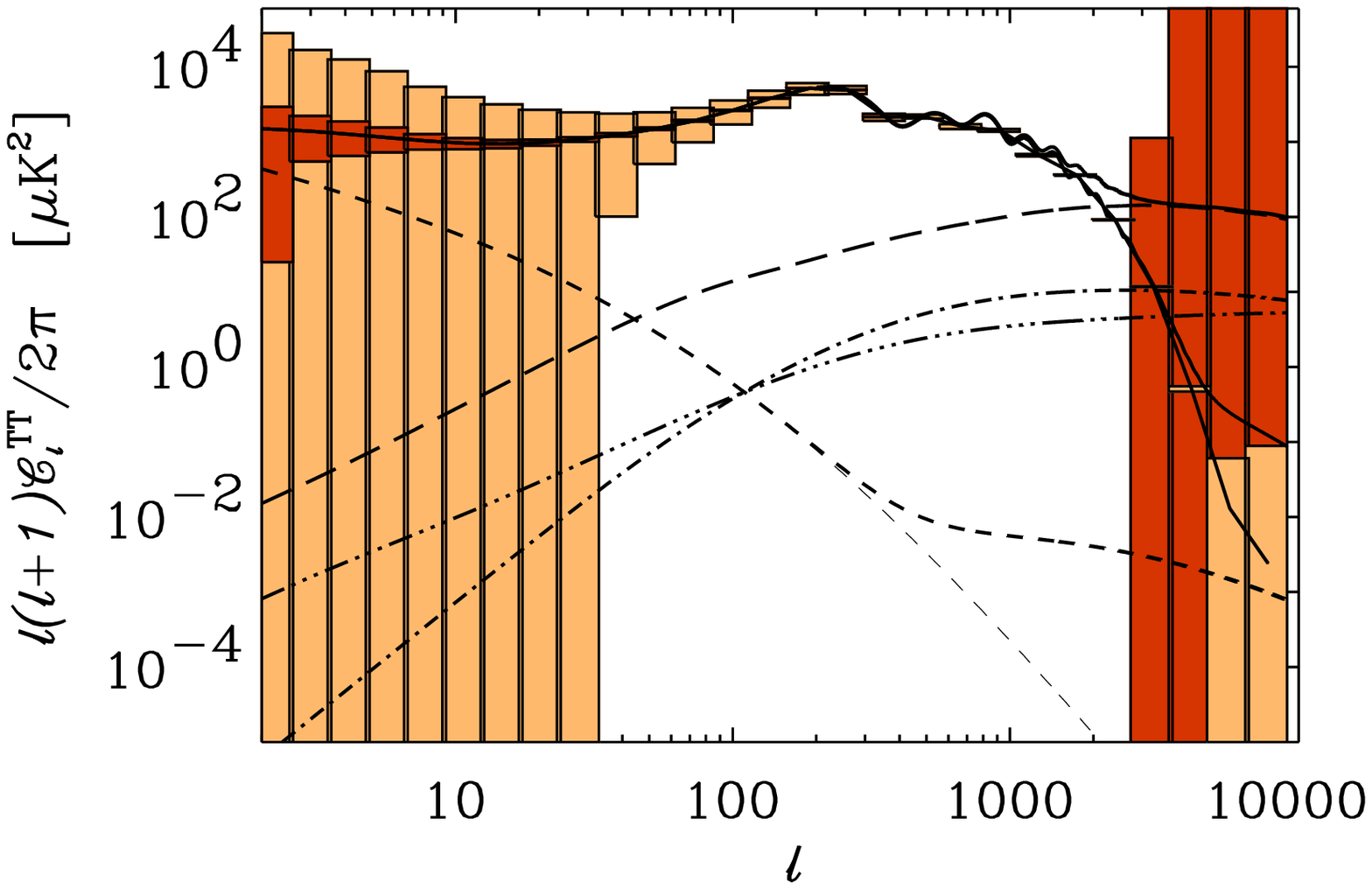,width=9.5cm}\epsfig{file =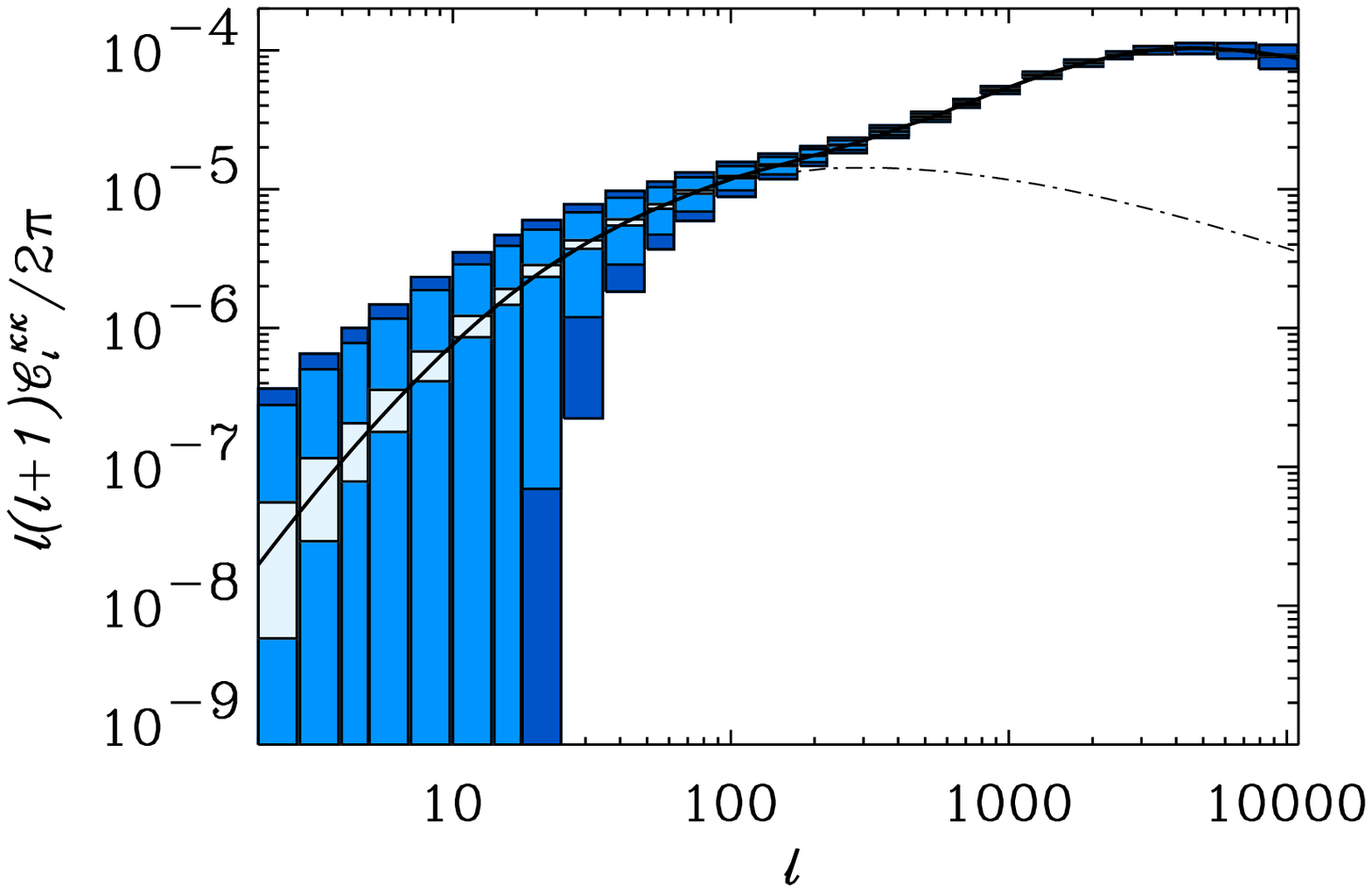,width=9.5cm}}
  \caption{ \emph{Left:} Expected signal and errors for the coming PLANCK and ACT
  CMB experiments (see \S\ref{sec:prospects} for precise assumptions). The
  Bordeaux wine color refers to PLANCK whereas the cream one refers to
  ACT. The cosmic variance is computed assuming the primordial power
  spectrum only, which is obviously an lower limit. The thick solid
  line denotes the primordial CMB signal with (up) or without (down) the lensing
  contribution. The thick short-dashed line denotes the ISW effect (low
  $\ell$) and the non-linear RS effect (high $\ell$). Note the angular
  transition between both. The thick long-dashed line denotes the
  expected thermal SZ contribution according to \citep{KoSe02}. The
  thick double-dot-dashed line corresponding to the expected kSZ,
  whereas the dot-dashed line corresponds to a possible component
  induced by patchy-reionisation (note that the overall amplitude is
  rather model-dependent and that we here assume the strongest signal
  \citep{Sa03}). The thin solid line denotes the sum of all those contributions.
\emph{Right:}Expected signal and error for the coming CFHTLS, SNAP and
  LSST experiments (see \S\ref{sec:prospects} for precise
  assumptions). We consider here only shot noise due to intrinsic
  ellipticities and cosmic variance, and assume that both dominates
  over systematics at any scale. We assume for this illustration only that this 3
  surveys measure the same convergence power spectrum. We thus
  neglect the effects of the different sources populations. From
  darker to lighter, the 3 depicted error boxes correspond to CFHTLS,
  SNAP and LSST.}
\label{fig:plot_clt}
\end{figure*}

The overlap of angular scales between secondary anisotropies of the CMB and 
cosmic shear is illustrated in figure~\ref{fig:plot_clt}. Error bars are shown 
for current and future CMB experiments and weak lensing surveys.  
Theoretical estimates for primary and secondary anisotropies are shown in 
the left panel, and predictions for the power spectrum of the weak 
lensing convergence field is shown at right.  \citet{Se01} calculated the 
correlation between cosmic shear and the thermal Sunyaev-Zel'dovich 
effect (tSZ).  Once the tSZ signal is removed by frequency cleaning, 
the Rees-Sciama effect, CMB lensing, kinetic SZ, and possibly patchy 
reionization remain. \citet{Hu02} considered the correlation between 
cosmic shear and both the linear ISW effect and  CMB lensing.  
As is visible in the figure,  
the nonlinear correction to the ISW effect, known as the Rees-Sciama effect 
is important on small scales $\ell \gtrsim 300$, and we compute the 
correlation between cosmic shear and the Rees-Sciama effect in a 
companion paper \citep{DoHe03}. For $\ell \gtrsim 3000$, the kinetic 
SZ spectrum  intersects the lensed CMB damping tail, and will likely 
dominate the anisotropy spectrum, (we have plotted the maximal patchy 
reionization signal for illustration). 

Below we calculate the fully nonlinear cross correlation between cosmic 
shear and the kinetic SZ effect. This calculation is 
complicated by the fact that the kSZ effect is proportional to the velocity 
of the ionized baryons, $\Delta T/T \sim \vecv$; hence, any two point 
correlation with a density tracer will be negligible for an isotropic 
velocity field. This is simply the
statement that the over-density field $\delta$ is just as likely to correlate
with a cluster moving toward us as one moving away from us, and thus
the average correlation vanishes. In fact, the isotropy of the velocity field 
guarantees that any statistic of odd powers of the velocity field is highly 
suppressed relative to even statistics (Monin \& Yaglom 1971,
Scannapieco 2000, Castro 2003). Thus, we must work with a three point
statistic -- two kSZ points and one weak lensing --- and we consider a ``collapsed'' 
configuration. That is we calculate the two point function of weak lensing 
and the filtered CMB temperature squared, which condenses three point 
information into an easily measurable power spectrum. 

We  model the observables in \S~\ref{sec:model} and calculate the angular 
auto power spectra of weak lensing and kinetic SZ which are shown
fig.~\ref{fig:plot_clt}. Special attention is paid to the approximations 
and analytical techniques used to calculate these auto power spectra in the 
fully nonlinear regime. The cross-correlation signals are calculated in  
\S~\ref{sec:cross-corr}. In \S~\ref{sec:kernel} we attempt to qualitatively 
understand why cosmic shear is correlated with the square of kinetic SZ, and 
the dependence of the correlation signal on the power spectrum normalization 
is explored in \S~\ref{sec:sig8}.  
A signal to noise ratio analysis is performed in  \S~\ref{sec:prospects},
to quantify the prospect for detecting this correlation  for current
and future experiments. We discuss our results, discuss limitations,
and conclude in \S~\ref{sec:discuss}. 

Throughout this paper we consider only flat cosmologies.  As our fiducial
model we choose the best fit WMAP (only) model of \citet{Sp03}, with
 $\Omega_m = 0.270$, $\Omega_\Lambda =0.73$, $h=0.72$, $n_s=0.97$,
$\sigma_8=0.84$ and $\tau = 0.17$.

\section{Modeling the Observables}
\label{sec:model}

We define in this section the needed general equations before
introducing the effect that will be discussed throughout this paper.
We calculate their angular auto power spectra and sketches their cross-correlation.
The calculations in this section are complicated by the fact that the
three dimensional fields whose power spectra and bispectra we desire are
highly nonlinear. Special attention is paid to the approximations and 
analytical techniques used to calculate these nonlinear power spectra.

\subsection{Angular Power Spectra in the Flat Sky Approximation}

We will be interested in the angular power spectra of fields on the sky 
$X(\hatn)$ which are weighted line-of-sight projections of three 
dimensional fields which we denote $\delta_X$, 
\be
X(\hatn) = f^X \star \Bigl( \int d\eta \ W^X(\eta)\delta_X(\eta\hatn,\eta)\Bigr) \label{eqn:Xdef}
\ee 
where $W^X(\eta)$ is the weight function for the field $X$ and $f^X$ is
a real space isotropic filter with which we convolve this field. We define
the conformal ``look back'' time as $\eta = \tau_0 - \tau$, which can also be
interpreted as the comoving distance. Here $\tau$ is the conformal time 
$d\tau = (1/a)$, $\tau_0$, which can also be interpreted as the comoving 
distance from the observer. Here and throughout we set $c=1$

For small sections of the sky or high multipole moments, it is a good 
approximation to treat the sky as flat. In this flat-sky approximation, the 
Fourier moments of this field on the sky are
\be
\tilde X (\bfell) = \int d^2\vecn \ X(\hatn) e^{-i\bfell\cdot\hatn} 
\label{eqn:Fourier}
\ee
Combining with eq.~(\ref{eqn:Xdef}) gives
\be
\tilde X (\bfell) = f^X(\ell) \int \frac{d\eta}{\eta^2}W^X(\eta)\int
\frac{dk_z}{2\pi}\tilde\delta(\veck_{\perp}=\bfell/\eta,k_z)e^{ik_z\eta},
\label{eqn:X_l}
\ee
where $f^X(\ell)$ is the Fourier transform of $f^X$. 

As in the all sky case, the cross correlation (or auto power spectrum) for any 
two fields, $X$ and $Z$ is defined by
\be
\langle \tilde X^{\ast}(\bfell)\tilde Z(\bfell)\rangle=(2\pi)^2\delta_{D}(\bfell+\bfell^{
\prime})C^{XZ}(\ell), \label{eqn:Cl_def}
\ee
and similarly the power spectrum between two three dimensional fields 
$\delta_X$ and $\delta_{Z}$ is
\be
\langle \tilde \delta_X({\bf k},\eta) \tilde \delta_Z({\bf k},\eta)\rangle 
= (2\pi)^3\delta_D({\bf k}+{\bf k'})P_{XZ}(k,\eta) ,  \label{eqn:ps_def}
\ee
Where $\delta_D$ is the Dirac delta function. 

For the relationship between these flat-sky Fourier modes and the 
spherical harmonic coefficients $X_{\ell m}$ and a proof of the
correspondence between the angular power spectrum $C^{XZ}_{\ell}$ and 
$C^{XZ}(\ell)$, see e.g. \citet{White99} or Appendix C of \citet{Hu00b}.

Combining eqs.~(\ref{eqn:X_l}), (\ref{eqn:Cl_def}), and (\ref{eqn:ps_def}), 
the correlation between $X$ and $Z$ is 
\bea
\langle \tilde X^{\ast}(\bfell)\tilde Z(\bfell^\prime)\rangle &=& (2\pi)^2 
f^X(\ell)f^Z(\ell)\int d\eta \frac{W^X(\eta)}{\eta^2}\nonumber\\
&& \times \int d\eta^\prime\frac{W^Z(\eta^\prime)}{\eta^{\prime 2}}\int \frac{dk_z}{2\pi}e^{i k_{z}(\eta-\eta^{\prime})} \nonumber\\
&& \delta_{D}(\bfell/\eta+\bfell^\prime/\eta^{\prime})
P_{XZ}(\sqrt{(\bfell/\eta)^{2}+k^2_z})\label{eqn:XZ_full}, 
\eea

Provided that the window functions $W(\eta)$ are slowly varying, 
$k \gg {\dot W^X}/W^X$, we can work in the Limber approximation \citep{Ka92}, 
allowing us to neglect radial modes relative to perpendicular modes, 
$k_z \ll \ell/\eta$, and we arrive at
\bea
\Delta^2_{XZ}(\ell) &=&\frac{\pi}{\ell}f^X(\ell)f^Z(\ell)\int d\ln\eta \ 
\eta^2 W^X(\eta)W^Z(\eta)\nonumber\\
&& \quad \times \Delta^2_{XZ}(k=\ell/\eta,\eta) \label{eqn:cross_co}
\eea
where the we have introduced the dimensionless angular power spectrum 
$\Delta^2_{XZ}(\ell)=\frac{\ell^2}{2\pi}C^{XZ}_{\ell}$, and the 
dimensionless 3-d power spectrum 
$\Delta^2_{XZ}(k,\eta)=k^3/2\pi^2 P_{XZ}(k,\eta)$. In the following section, 
we list the relevant window functions and discuss the phenomenological 
techniques used to calculate the three dimensional power spectra in the fully
nonlinear regime. Note that until \S~\ref{sec:prospects}, we set $f^X(\ell) = 1$. 

\subsection{Weak Gravitational Lensing}

The distortion of background source galaxies by the foreground matter 
distribution is completely described by the convergence field $\kappa(\hatn)$ 
in the weak lensing approximation, which is a weighted projection of the 
density field along the line of sight
\be
\kappa(\hatn) = \int_0^{\tau_0}d\eta W^{\kappa}(\eta)\ \delta(\eta\hatn,\eta) ,
\ee
where the kernel is given by (for a flat cosmology)	
\be
W^{\kappa}(\eta) = \frac{3\Omega_{m0}H_0^2\eta}{2 a}\int_\eta^{\tau_0}d\eta' {\eta'-\eta \over \eta'}S(\eta')\; ,
\ee
and $S(\eta')$ describes the normalized radial distribution of
sources
\be
S(\eta)=p_z(z)\frac{d\eta}{dz}, 
\ee
with $p_z(z)$ being the normalized source redshift distribution 
$\int dz p_z(z) = 1$, which in principle can be measured from the weak lensing
survey. For definiteness, we use
\beq
p_z\left(z\right)=\frac{1}{2z_0^3}z^2 e^{-z/z0}. \label{p_z}
\eeq
This redshift distribution peaks at $2z_0$, has mean 
redshift $\langle z \rangle=3z_0$, and has been used in previous studies 
of cosmic shear (Wittman et al. 2000). The values of $z_0$ that we 
use to represent current and future surveys, along with the other 
specifications are listed in table 2. 

In order to evaluate the weak lensing power spectrum we need an expression
for the fully nonlinear power spectrum of the over-density field $P_{\delta\delta}(k)$.  Fitting formula for the nonlinear power spectrum have been 
studied extensively \citep{Ham91,PeDo96,Ma98,Smith03}. We use the 
scaling formula of \citet{Smith03} for the nonlinear power spectrum, 
which is accurate at better than the $10\%$ level.

The angular power spectrum $C^{\kappa}_{\ell}$ of the \emph{cosmic shear} is
shown in fig.~\ref{fig:plot_clt}.  Error bars are shown for current and 
future surveys which will have different source redshift distributions, 
although for illustration, we show only the case $z_0=0.5$.

\subsection{The kinetic SZ effect}

CMB photons diffusing to us from the surface of last scattering are  
up-scattered by the electron plasma bound in the gravitational potentials
of dark matter halos. The motion and density variations of those
scatterers will imprint a specific thermal temperature fluctuation pattern 
on the CMB, $\Theta^{\rm kSZ}=\Delta T_{kSZ}/T_0$, both because of the 
Doppler effect and of variations in opacity \citep{sunyaev80},
\be
\Theta^{\rm kSZ}\left(\hatn\right)=\int_0^{\tau_0} d\eta\ g\left(\eta\right) \hatn\cdot\vecp(\hatn\eta,\eta) 
\ee
where we have introduced the momentum density
\be
\vecp(\hatn\eta,\eta) =(1+\delta(\hatn\eta,\eta)\bigr)\vecv(\hatn\eta,\eta)\;
\ee
and the visibility function, $g$. The latter can be expressed in terms of the
Thomson optical depth, $\tau(\eta)=$$\int_0^{\eta}\dot\tau(\eta)d\eta$, where
$\dot\tau(\eta)=a(\eta)n_ex_e\sigma_T$ with $\sigma_T$ the Thomson
scattering cross section, $n_e$ the electron density and $x_e$ the
ionization fraction. Thus
\be
g\left(\eta\right)=\frac{d\tau}{d\eta}e^{-\tau} = x_e \tau_{H}\left(1+z\right)^2 e^{-\tau}\; ,
\ee 
where
\be 
\tau_H = 0.0691\left(1-Y_p\right)\Omega_b h ,
\ee 
is the Thomson scattering optical depth to the Hubble distance today, 
assuming full hydrogen ionization, and $Y_p$ is the primordial helium fraction.
We assume further that reionization occurs instantaneously, \ie $x_e=1$ at all
times after the reionization epoch, $z_{ri}$. Then we have \citep{Gr99}
\be
\tau(z_{ri}) = \frac{2}{3} \frac{\tau_H}{\Omega_m}\left[\sqrt{1-\Omega_m + \Omega_m\left(1+z_{ri}\right)^3} -1\right]
\ee
for a flat universe.

As was first pointed out by \citet{Ka84}, cancellation of successive peaks 
and troughs for small wavelength modes prevents a significant kinetic SZ 
effect from contributing to the CMB anisotropy at linear order. This can be 
understood in the context of the 'Limber' \citep{Ka92} or 
'weak coupling' \citep{HuWh96} approximations. They dictate that only modes 
with wave-vectors perpendicular to the line of sight contribute to the 
projection of a given random field, because uncorrelated
radial waves cancel in the line of sight projection.  Gravity generates 
potential  velocity flows, so that radial velocities have no wave-vector 
component perpendicular, in Fourier space, to the line of sight; 
consequently, the linear Doppler effect $\vecp \sim \vecv$ vanishes.  

\citet{OsVi86} showed that the same will be not be true for density
modulated velocity flows, $\vecp \sim \delta \vecv$, which can have a
non-vanishing curl. Thus wave-vectors perpendicular to the line  
of sight generate fluctuations parallel to the line of sight. 
At second order in perturbation theory, this is known as the 
Ostriker-Vishniac effect \citep{OsVi86,Vi87,DoJu95,HuWh96}; whereas, the full
nonlinear signal is referred to as the kinetic-SZ (kSZ) effect. This 
full non-linear contribution  has been studied in detail both analytically 
\citep{Hu00a,MaFr01,Zh03} and numerically 
\citep{GnJa01,Sp01,daS01,Zh03}. 

Although the amplitude of the expected signal is uncertain, the maximal 
signal can be calculated if we assume that the electrons trace the dark matter
exactly. Then we can write for the power spectrum of the radial 
momentum component $p_{\hatn}$ \citep{Hu00a,MaFr00}
\be 
P_{p_{\hatn}p_{\hatn}}\left(k\right)\approx \frac{1}{2}P_{\vecp_{\perp}\vecp_{\perp}}\left(k\right) \approx
\frac{1}{3}v^2_{\rm rms}P^{\rm{nl}}_{\delta\delta}. \label{MaFrya} 
\ee
where $P_{\vecp_{\perp}\vecp_{\perp}}$ is the power spectrum of the 
\emph{vortical} component of the momentum field $\vecp_{\perp}$. 
This assumes the dominant contribution to the power spectrum of the radial 
momentum comes from nonlinear densities coupling to linear bulk velocity flows
(Hu 2000; Ma \& Fry 2001; though see also Zhang \etal 2003). The right panel 
of fig.~\ref{fig:hkern_KSZ} shows the 3-d auto power spectrum 
$P_{p_{\hatn}p_{\hatn}}$ at $z=0.5$. The projection of this power spectrum 
with eq.~(\ref{eqn:cross_co}) gives the angular kSZ power spectrum, which is 
shown in fig.~\ref{fig:plot_clt}.

\section{The kinetic SZ weak lensing correlation}
\label{sec:cross-corr}

In this section we compute the correlation between the kinetic SZ effect and 
weak gravitational lensing. This calculation is complicated by the fact 
that the simple two point correlation 
$\langle \kappa \Theta\rangle$, will vanish because 
$\Theta^{\rm kSZ}\sim \vecv$, and the velocity field is isotropic.  
This is just the statement that the over-density field 
$\delta$ is just as likely to correlate with a cluster moving toward us as one
moving away from us, and the average correlation vanishes. The isotropy of 
the velocity field  guarantees that statistics of odd powers of the velocity 
field are highly suppressed relative to even 
statistics (Monin \& Yaglom 1971, Scannapieco 2000, Castro 2003).

Thus a non-vanishing correlation requires that we work with a three point 
statistic: two kSZ points, so that our statistic is even in the velocity,
and one point weak lensing. We consider the simplest case of a ``collapsed''
three point function --- the cross correlation between the kappa field and 
the square of the kSZ, $\langle\kappa \Theta^{2}\rangle$ --- which 
condenses three point information into an easily measurable angular 
power spectrum. 

Squaring the temperature field couples multipoles at all scales, and in 
particular, power from multipoles outside the range where the kSZ is 
dominant, will swamp the signal.  To prevent our quadratic temperature 
statistic from being polluted by the beam and primary anisotropies, 
we must filter the temperature before squaring
\be
\tilde\Theta_{f}(\bfell)=f(\bfell)\tilde\Theta(\bfell)
\ee
This simple convolution before squaring is a special case of a more 
general class of filters for quadratic temperature statistics studied by 
\citet{Hu02}.

\subsection{Limber approximation for the bispectrum}

In the flat sky approximation, the square of the filtered temperature
fluctuation can be written as a convolution in (2D) Fourier space,
\be
\tilde\Theta_f^2(\bfell^\prime) = \int \frac{d^2\bfell^{\prime\prime}}{\left(2\pi\right)^2}
\tilde\Theta_f(\bfell^{\prime\prime}) \tilde\Theta_f(\bfell^\prime-\bfell^{\prime\prime}),  \label{conv_theta}
\ee
and the weak lensing kinetic SZ squared correlation is
\be
\langle \tilde\kappa(\bfell)\tilde\Theta_f^2(\bfell^\prime)\rangle=\int
\frac{d^2\bfell^{\prime\prime}}{\left(2\pi\right)^2} \langle
\tilde\kappa(\bfell)\tilde\Theta_f(\bfell^{\prime\prime})\tilde
\Theta_f(\bfell^\prime-\bfell^{\prime\prime})\rangle
\label{eqn:kappa_KSZ2}  
\ee
with
\bea
&&\langle\tilde\kappa(\bfell) \tilde\Theta_f(\bfell^{\prime\prime})\tilde
\Theta_f(\bfell^\prime-\bfell^{\prime\prime})\rangle
= (2\pi)^2 f(\ell)f(|\bfell^{\prime}-\bfell^{\prime\prime}|)\nonumber\\
&& \times \int d\eta \frac{W^{\kappa}\left(\eta\right)}{\eta^2}
\int d\eta^\prime 
\frac{g\left(\eta^\prime\right)}{\eta^{\prime 2}}\int d\eta^{\prime\prime}
\frac{g\left(\eta^{\prime\prime}\right)}{\eta^{\prime\prime 2}}\nonumber\\
&& \times \int \frac{dk_z}{2\pi} e^{i k_{z}(\eta-\eta^{\prime})}\int 
\frac{dk^{\prime}_z}{2\pi}e^{ik^{\prime\prime}_{z}(\eta^{\prime\prime}-\eta^{\prime})}\\
&& \times  \ \delta_{D}(\bfell/\eta + \bfell^{\prime\prime}/\eta^{\prime
\prime}+(\bfell^{\prime}-\bfell^{\prime\prime})/\eta^{\prime}) \nonumber\\
&& \times \ B_{\delta
p_{\hatn}p_{\hatn}}(\bfell/\eta +\veck_{z},\bfell^{\prime\prime}/\eta^{\prime\prime}+\veck^{\prime\prime}_z,(\bfell^{\prime}-\bfell^{\prime\prime})/\eta^{\prime}-(\veck_z+\veck^{\prime\prime}_z))\nonumber,\label{eqn:big_mess}
\eea
where we have introduced the \emph{hybrid} bispectrum 
\bea
\langle\tilde\delta&&\left(\veck_1\right)\tilde
p_{\hatn}\left(\veck_2\right)\tilde p_{\hatn}\left(\veck_3\right)\rangle = \nonumber\\
  && \left(2\pi\right)^3 B_{\delta
  p_{\hatn}p_{\hatn}}\left(\veck_1,\veck_2,\veck_3\right)\delta_D\left(\veck_1 + \veck_2 + \veck_3\right). \label{bispect} 
\eea

The expression in eq.~(\ref{eqn:big_mess}) can be simplified in the 
Limber approximation for the bispectrum \citep{BuKa00}, valid for the 
small angles considered here. We can again ignore radial modes 
$k_z \ll \ell/\eta$, giving 
\bea
\langle
&& \tilde\kappa(\bfell)\tilde\Theta_f(\bfell^{\prime\prime})\tilde\Theta_f(\bfell^\prime-\bfell^{\prime\prime})\rangle\nonumber
\approx  
 (2\pi)^2 \delta_D(\bfell + \bfell^{\prime})f(\ell^{\prime\prime})f(|\bfell^{\prime}-\bfell^{\prime\prime}|)\nonumber\\
  && \times \int \frac{d\eta}{\eta^4}W^\kappa\left(\eta\right)\left[g\left(\eta\right)\right]^2 B_{\delta p_{\hatn}p_{\hatn}}\left(\bfell/\eta,\bfell^{\prime\prime}/\eta,(\bfell^{\prime}-\bfell^{\prime\prime})/\eta\right) 
\eea
Plugging this into eq.~(\ref{eqn:kappa_KSZ2}) gives 
\be
\langle \kappa(\bfell)\Theta_f^2(\bfell^\prime)\rangle = \left(2\pi\right)^2
\delta_D\left(\bfell + \bfell^{\prime}\right) C^{\kappa\Theta_f^2}(\ell)
\ee
with
\be
C^{\kappa\Theta_{f}^2}(\ell)= \int \frac{d\eta}{\eta^2}W^\kappa\left(\eta\right)\left[g\left(\eta\right)\right]^2 {\cal T}(k=\ell/\eta,\eta), \label{eqn:Cl}
\ee
and where we have defined
\be
{\cal T}(k,\eta) \equiv \int \frac{d^2\vecq}{\left(2\pi\right)^2}  \  
f(q\eta)f(|\veck+\vecq|\eta)B_{\delta p_{\hatn}p_{\hatn}}\left(\veck,\vecq,-\veck-\vecq\right) \ .\label{eqn:Tofk} 
\ee

This power spectrum in eq.~(\ref{eqn:Cl}) can be thought of as a 
Limber projection of the ``power spectrum'' ${\cal T}(k,\eta)$ which 
is an integral over all \emph{planar} triangle configurations
of the bispectrum $\left(\veck,\vecq,-\veck-\vecq\right)$, with
one of the sides of length $k$. It depends only on the norm of $\veck$
by statistical isotropy and effectively collapses the three point
information contained in $B_{\delta p_{\hatn}p_{\hatn}}$
\citep{Co01b}. We henceforth refer to eq.~(\ref{eqn:Tofk}) as the \emph{triangle power spectrum}.   

\subsection{The hybrid bispectrum $B_{\delta p_{\hatn}p_{\hatn}}$}

In order to calculate the weak lensing kinetic SZ squared correlation, we must
evaluate the hybrid bispectrum $B_{\delta p_{\hatn}p_{\hatn}}$ of the density
and radial momentum component.  Here we proceed by analogy with previous 
analytical studies of the kSZ power spectrum. The power spectrum of the radial momentum component is
\be
\langle
p_{\hatn}\left(\veck\right)p_{\hatn}\left(\veck^{\prime}\right)\rangle
= \left(2\pi\right)^3\delta_D\left(\veck + \veck^{\prime}\right)P_{p_{\hatn}p_{\hatn}}\left(k\right)\; .
\ee
\citet{Hu00a} and then the joint numerical and analytical work of
\citet{MaFr00} suggested that the dominant contribution comes from
non-linear densities coupling to linear bulk velocity
flows. Furthermore, because of the cancellations that occur in the
projection \citep{Ka84}, only the curl component of the projected
momentum $\vecp$ - \ie the component perpendicular to $\hatn$ in the Fourier
domain that we thus note $\vecp_{\perp}$- will contribute to the kSZ temperature
fluctuation, so that 
\be 
P_{p_{\hatn}p_{\hatn}}\left(k\right)\approx \frac{1}{2}P_{\vecp_{\perp}\vecp_{\perp}}\left(k\right) \approx
\frac{1}{3}v^2_{\rm rms}P^{\rm{nl}}_{\delta\delta}. \label{MaFry} 
\ee
where $v^2_{\rm rms}$ is the volume averaged velocity dispersion
\be
v^2_{\rm rms} = \int \frac{d^3\veck}{\left(2\pi\right)^3} P_{vv}\left(k\right).
\ee

We follow this analytical line of reasoning and make the following ansatz
\be
B_{\delta p_{\hatn}p_{\hatn}} \approx \frac{1}{3}v^2_{\rm rms}B^{\rm{nl}}_{\delta\delta\delta}\ ,
\ee
which presumes that the dominant contribution to the bispectrum in 
eq.~(\ref{bispect}) comes from large scale bulk velocity flows coupling to 
the three point function of the nonlinear density field. Note that at the 2
points level, analogous approximations can be motivated in a halo model context
\citep{CoSh02,Sh03}. Note however that it was also claimed, even if not clearly
illustrated, that the curl component of the momentum induced by
the vorticity in the non-linear velocity field itself might not be
completely negligible \citep{Zh03}. If this were of some importance,
then our evaluation would \emph{underestimate} the signal of interest.
 
We are interested in the high-$k$, non-linear behavior of ${\cal T}\left(k\right)$,
thus we need to evaluate the integral in eq.~(\ref{eqn:Tofk}) with the fully 
nonlinear bispectrum of the density field. We use for this purpose the fitting function
of \citet{ScCo01} for the bispectrum, \ie 
\be
B^{\rm{nl}}_{\delta\delta\delta} =
2\Feff\left(\veck_1,\veck_2\right)P^{\rm
  nl}_{\delta\delta}\left(k_1\right)P^{\rm
  nl}_{\delta\delta}\left(k_2\right) + {\rm cyclic} \label{scocci} 
\ee
where
\bea
\Feff\left(\veck_1,\veck_2\right)  = && \frac{5}{7}a\left(\neff,k_1\right) a\left(n_{\rm eff},k_2\right)\nonumber \\ 
+ \quad\frac{1}{2}\frac{\veck_1\cdot\veck_2}{k_1 k_2}
\left(\frac{k_1}{k_2} \right. +&& \left. \frac{k_2}{k_1}\right)b\left(\neff,k_1\right)b\left(\neff,k_2\right) \nonumber\\
+ \quad \frac{2}{7}\left(\frac{\veck_1\cdot\veck_2}{k_1 k_2}\right)^2&&c\left(\neff,k_1\right)c\left(\neff,k_2\right)\ .\label{Feff}
\eea
The fitting functions are given by
\bea
a\left(\neff,k\right) & = & \frac{1+\sigma_8^{-0.2}\left(z\right)\sqrt{0.7 \ Q_3\left(\neff\right)}\left(q/4\right)^{\neff + 3.5}}{1 + \left(q/4\right)^{\neff + 3.5}}, \nonumber\\
b\left(\neff,k\right) & = & \frac{1 + 0.4\left(\neff + 3\right)q^{\neff + 3}}{ 1 + q^{\neff + 3.5}}, \\
c\left(\neff,k\right) & = & \frac{1+ \left(\frac{4.5}{1.5+\left(\neff
    + 3\right)^4}\right)\left(2q\right)^{\neff + 3}}{1 +
  \left(2q\right)^{\neff + 3.5}}\ , \nonumber \label{abc}
\eea
where $\neff$ is the effective spectral index of the power spectrum, 
\be
\neff\left(k\right) \equiv \frac{d \ln P}{d \ln k}\ .
\ee
The quantities $q$, $Q$, and $\sigma_8\left(z\right)$ are defined by
\be
q \equiv \frac{k}{k_{\rm nl}}    \ \ \ {\rm with} \ \ \  \frac{k^3}{2\pi^2}P^{\rm lin}_{\delta\delta}\left(k_{\rm nl},z\right) = 1 , 
\ee
\be
Q\left(\neff\right) = \frac{4-2^{\neff}}{1+ 2^{\neff + 1}}, 
\ee
and 
\be 
\sigma_8\left(z\right) = \int \frac{d^3\veck}{\left(2\pi\right)^3}P^{\rm lin}_{\delta\delta}\left(k,z\right)W\left(k R\right)
\ee
where $W(kR)$ is the usual Fourier transform of a top-hat of radius $R = 8\ \hMpc$. 

The functions $a\left(\neff,k\right)$,  $b\left(\neff,k\right)$, and 
$c\left(\neff,k\right)$ in eq.~(\ref{abc}) interpolate between the one loop 
perturbative expansion and highly non-linear regimes for general CDM
cosmological models. 
It can be seen that for large scales, \ie $k \ll k_{\rm nl} $, $a=b=c=1$ and the
tree level perturbation theory expression is recovered, whereas on
small scales, \ie $k \gg k_{\rm nl}$, $ a = \sigma_8^{-0.2}$ $\left(z\right)\sqrt{0.7\ Q_3\left(\neff\right)}$ and
$b=c=0$, so that the bispectrum becomes independent of triangle configuration. 

\subsection{The Triangle Power Spectrum}

In this section we evaluate the triangle power spectrum for the simplest 
case where the filter functions are set to unity, $f(\ell)=1$. In this case,
a simple approximation exists in the high $k$ limit that dramatically 
simplifies the computation and provides and intuitive understanding of
the source and strength of the correlation.  The full computation will be 
performed in for the purpose of computing the signal to noise ration 
in \S \ref{sec:prospects}, one the appropriate filter is introduced.

The integral in eq.~(\ref{eqn:Tofk}) will have three terms corresponding to 
the three permutations of the wave vectors in eq.~(\ref{scocci})
\be
{\cal T}\left(k\right)=\frac{1}{3}v^2_{\rm rms}\left[ {\cal
    T}_1\left(k\right) + {\cal T}_2\left(k\right) + {\cal T}_3\left(k\right)\right] 
\ee
where
\bea
{\cal T}_1\left(k\right) & = & 2 P\left(k\right) \int \frac{d^2\vecq}{\left(2\pi\right)^2} \ \Feff\left(\veck,\vecq\right)P^{nl}_{\delta\delta}\left(q\right) \qquad \label{T1} \\ 
{\cal T}_2\left(k\right) & = & 2 P\left(k\right) \int \frac{d^2\vecq}{\left(2\pi\right)^2} \ \Feff\left(\veck,-\veck-\vecq\right)P^{nl}_{\delta\delta}\left(\norm\right) \label{T2}
\nonumber\\ 
{\cal T}_3\left(k\right) & = & 2 \int \frac{d^2\vecq}{\left(2\pi\right)^2}  \ \Feff\left(\vecq,-\veck-\vecq\right)P^{nl}_{\delta\delta}\left(q\right)P^{nl}_{\delta\delta}\left(\norm\right)\; .
\nonumber\label{T3} 
\eea

We first remark that ${\cal T}_2={\cal T}_1$ after a simple translation of the integration 
variable.  Then, ${\cal T}_1$ can be evaluated exactly in polar coordinates
because the angular integration factorizes out of the radial
integration  
\bea
{\cal T}_1\left(k\right)  & = & \frac{P^{\rm
  nl}_{\delta\delta}\left(k\right)}{7\pi}\left[5 \ a\left(k\right) \int dq \
  q \ a\left(q\right) P^{\rm nl}_{\delta\delta}\left(q\right)\right.\nonumber\\
  &&\quad \quad\quad +  \left. \quad c\left(k\right)\int dq \ q \
  c\left(q\right)P^{\rm nl}_{\delta\delta}\left(q\right)\right],
\eea
where we abbreviate $a\left(\neff,k\right)$ as $a\left(k\right)$, and
likewise for $b$ and $c$. The remaining term ${\cal T}_3$, can be written
\bea 
 {\cal T}_3\left(k\right) & = & 2 \int \frac{d^2\vecq}{\left(2\pi\right)^2}
 \left[\frac{5}{7}a\left(q\right)a\left(\norm\right)\qquad\qquad
   \right. \nonumber\\ 
 -       & & 
 \frac{1}{2}b\left(q\right)b\left(\norm\right)\left(\frac{q}{\norm} +
 \frac{\norm}{q}\right)\left(\frac{q}{k} + \mu\right) \nonumber \\
+& &
 \frac{2}{7}c\left(q\right)c\left(\norm\right)\left. \left(\frac{q}{k}
+ \mu\right)^2\right]P^{\rm nl}_{\delta\delta}\left(q\right)P^{\rm
   nl}_{\delta\delta}\left(\norm\right)\nonumber\\  
\label{T3b} 
\eea
with $\mu \equiv \hatk\cdot\hatq$.  

As mentioned previously, we are interested in the high-$k$, non-linear
behavior of ${\cal T}\left(k\right)$ as the weak lensing kinetic SZ squared
correlation will only be significant on small angular scales. The
dominant contributions to the integrals in eq.~(\ref{T3}) will be 
for $\vecq$ near the peak of $P^{nl}_{\delta\delta}\left(q\right)$ where 
$a\left(q\right) \sim b\left(q\right) \sim c\left(q\right) \sim 1$ . 
Thus in the high-$k$ limit the integrand is significant only for  $q \ll k$, 
and we can drop terms of order ${\cal O}\left(q/k\right)$. A similar approximation 
has been used previously to calculate the mode coupling integrals for the 
kinetic SZ power spectrum (Hu 2000a; Ma \& Fry 2001; Cooray 2001). 
In this approximation ${\cal T}_3 = {\cal T}_1$, and we get, introducing the
dimensionless power spectra, $\Delta^2_{\rm {\cal T}}$ and $\Delta^2_{\rm nl}$
\be
\Delta^2_{\rm {\cal T}}  = \frac{k^3}{2\pi^2}{\cal T}\quad {\rm and}\quad
\Delta^2_{\rm nl} = \frac{k^3}{2\pi^2}P^{\rm nl}_{\delta\delta}\
,\nonumber 
\ee
the following simple result 
\be
\Delta^2_{\rm {\cal T}}\left(k,z\right)= \frac{1}{3}v^2_{\rm
  rms}\left(z\right)\, \Delta^2_{\rm nl}\left(k,z\right)\,
E_3\left(k,z\right) \label{deltaT} 
\ee
where we defined the three point enhancement of the power spectrum
\bea
E_3\left(k,z\right) & = &\frac{6\pi}{7} \left[5 \ a\left(k,z\right)
  \int d\ln q \ a\left(q,z\right)\frac{\Delta^2_{\rm
      nl}\left(q,z\right)}{q}\right.\nonumber\\ 
  & + & \left. \quad c\left(k,z\right)\int d\ln q \
  c\left(q,z\right)\frac{\Delta^2_{\rm nl}\left(q,z\right)}{q}\
  \right]. 
\eea
Here the time dependence has been explicitly included in the functions above as a 
reminder. With these definitions, we obtain the final expression for
the weak-lensing kSZ squared power spectrum 
\be
\frac{\ell^2}{2\pi}C^{\kappa\Theta^2}_{\l} = \frac{\pi}{\ell}\int
d\eta \, \eta \,
W^\kappa\left(\eta\right)\left[g\left(\eta\right)\right]^2 \,
\Delta^2_{\rm {\cal T}}\left(k=\ell/\eta,\eta\right)\, . 
\ee

This intuitive result states that $C^{\kappa\Theta^2}_{\l}$ is given by a 
'Limber' projection of the triangle power spectrum, 
$\Delta^2_{\rm {\cal T}}\left(k,z\right)$, which condenses information from all 
planar triangle configurations of the hybrid density-momentum bispectrum with 
side length $k$. The form of the triangle power spectrum in eq.~(\ref{deltaT})
has as simple interpretation. It is similar to the kSZ power spectrum
in eq.~(\ref{MaFry}), in that it arises from density modulations of a
large scale bulk flow. However, here we are dealing with three point
modulations, which are \emph{enhanced} by the non-linear coupling of
small scale modes to large scale power. This mode coupling is
encapsulated in the integrals over the power spectrum in
$E_3\left(k,z\right)$.  The dimensionless triangle power spectrum 
eq.~(\ref{deltaT}) is shown in the right panel of fig.~(\ref{fig:hkern_KSZ}), 
along with the kSZ auto power spectrum eq.~(\ref{MaFry}). Note that they
have similar orders of magnitude.

\begin{figure*}[t]
\centering
\centerline{\epsfig{file=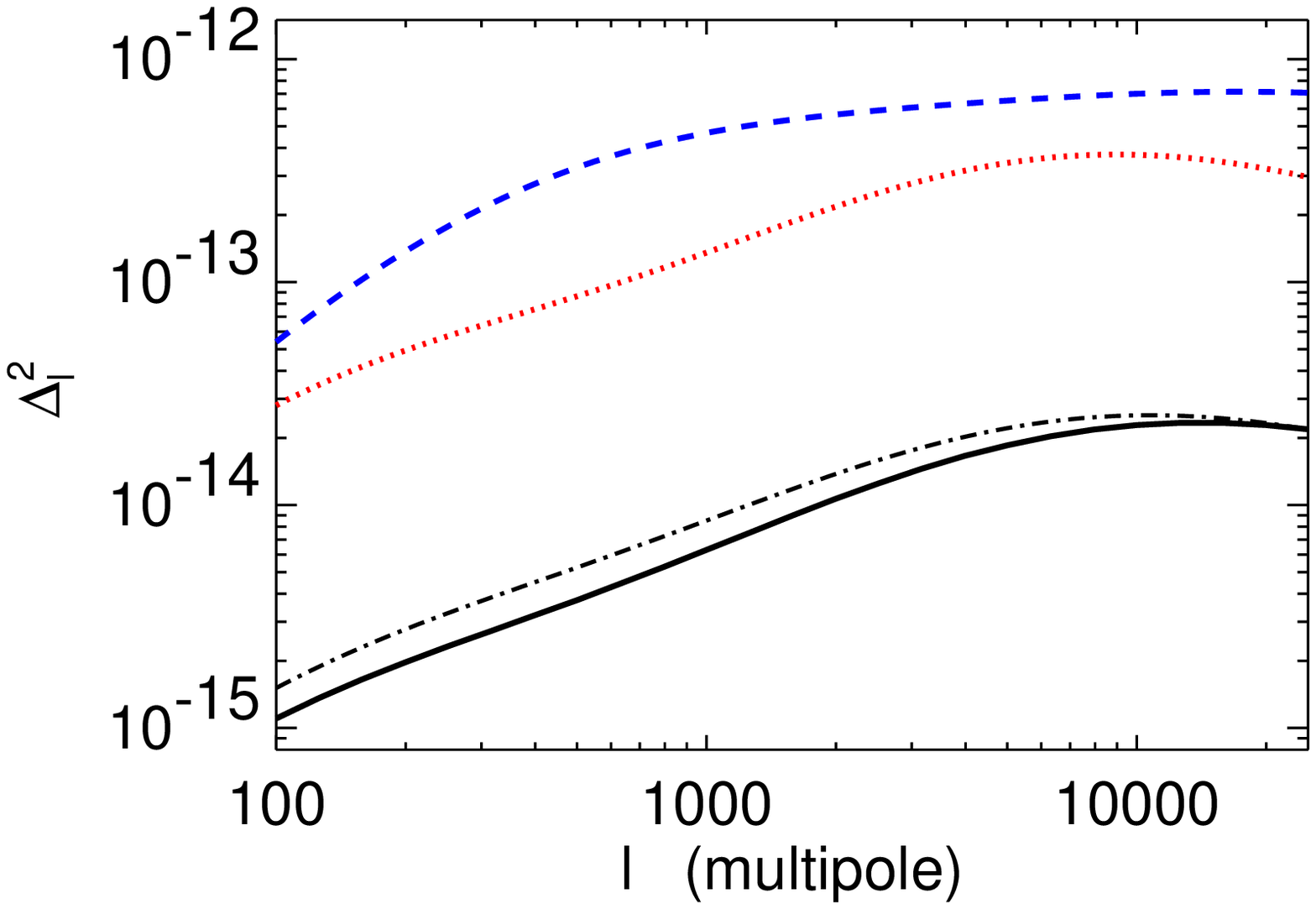,width=0.50\textwidth}\epsfig{file=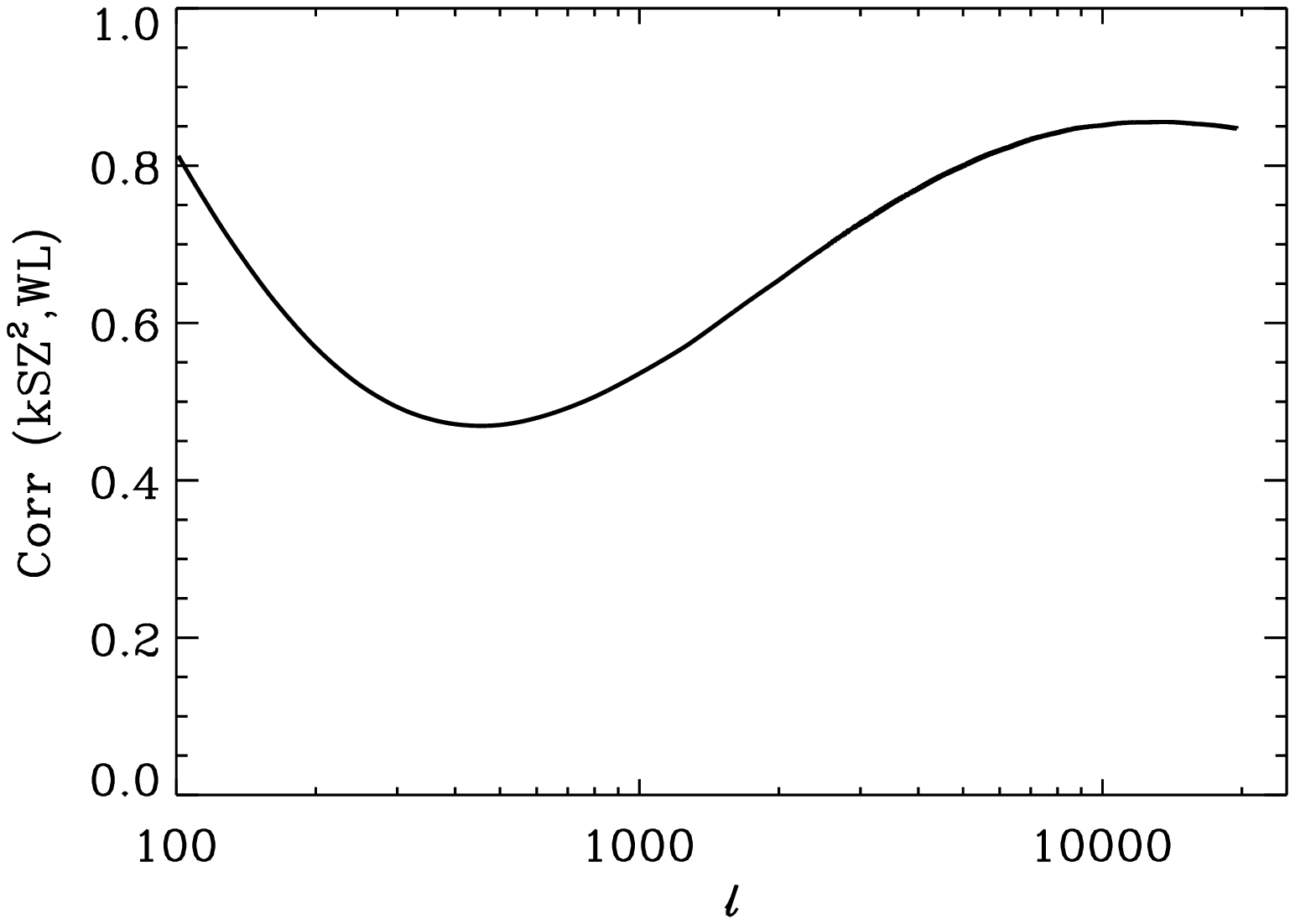,width=0.50\textwidth}}   
\caption{\emph{Left:}Angular power spectra for kinetic SZ and
  kSZ$^2$-weak lensing cross correlation. Only dimensionless units are
  used. The blue solid line corresponds to the kSZ$^2$ auto power
  spectrum, while the red solid line corresponds to the WL auto power
  spectrum scaled down by a factor $10^9$ for the sake of illustration. The kSZ$^2$-weak lensing
  angular cross power spectrum is shown by the solid (black) line. The
  dotted black line shows the result of using the approximation in
  eqn.~(\ref{deltaT}) for the triangle power spectrum, which is good
  to $\sim 25\%$ for $\ell \gtrsim 3000$.  The dashed (blue) line is
  the kinetic SZ auto power spectrum \emph{Right:} Cross correlation
  coefficient for the kSZ$^2$-weak lensing correlation.}\label{fig:d2l_kappa_KSZ2} 
\end{figure*}

\section{Understanding the Correlation}
\label{sec:kernel}

Before calculating the signal to noise for the aforementioned cross 
correlations, we attempt to understand the degree to which the secondary 
anisotropies are correlated with weak lensing.  

The strength of the angular correlation between any two random fields $X$ 
and $Y$ is quantified with the cross-correlation coefficient
\be
{\rm Corr}^{XY}(\ell) = \frac{C^{XY}_{\ell}}{\sqrt{C^{XX}_{\ell}
C^{YY}_{\ell}}}
\ee

As we are dealing with angular correlations of projected 
three dimensional fields (see eq.~(\ref{eqn:cross_co})), this cross-correlation 
coefficient will depend both on the extent to which the 
window functions $W^X(\eta)$ and $W^Y(\eta)$ overlap, and the strength of 
of the cross power spectrum $P_{XY}$ relative to the auto power spectra 
$P_{XX}$ and $P_{YY}$.

If we change the integration variables from $\eta$ to $k=\ell/\eta$, 
we can rewrite eq.~(\ref{eqn:cross_co}) for the angular power spectrum, 
as an integral over wavenumber
\be
\Delta^{2}_{XY}(\ell)=\int d\ln k \ H_{\ell}(k)\Delta^2_{XZ}(k,\eta=\ell/k) \label{cross_k}
\ee
where
\be
H^{XY}_{\ell}(k) \equiv \frac{\pi}{\ell}\left(\frac{\ell}{k}\right)^2 W^X(\ell/k)
 W^Y(\ell/k)
\label{eq:h_xy}
\ee
With the above two equations, we can think of any cross correlation as a 
weighted integral of the cross power spectrum, $\Delta^2_{XY}$, with the 
projection 'kernel' $H^{XY}_{\ell}(k)$. The kernels $H^{XX}_{\ell}(k)$ 
and $H^{YY}_{\ell}(k)$ indicate the scales probed by $X$ and $Y$ respectively, 
while the cross kernel $H^{XY}_{\ell}(k)$ indicates their degree of overlap.

The cross correlation coefficient for the kSZ$^2$-weak lensing correlation is 
plotted in fig.~{\ref{fig:hkern_KSZ}. Computing this required the auto power 
spectrum of the kSZ temperature fluctuation squared $C^{\rm kSZ^2}_{\ell}$ 
(see \S 5 eq.(~\ref{eq:clconvol})). The level of correlation is significant, 
approaching $\sim 0.8$ at the arc-min angular scales $\ell \gtrsim 3000$ in 
the damping tail, where the primary anisotropies are heavily attenuated. 

\begin{figure*}[t]
\centering
\centerline{\epsfig{file=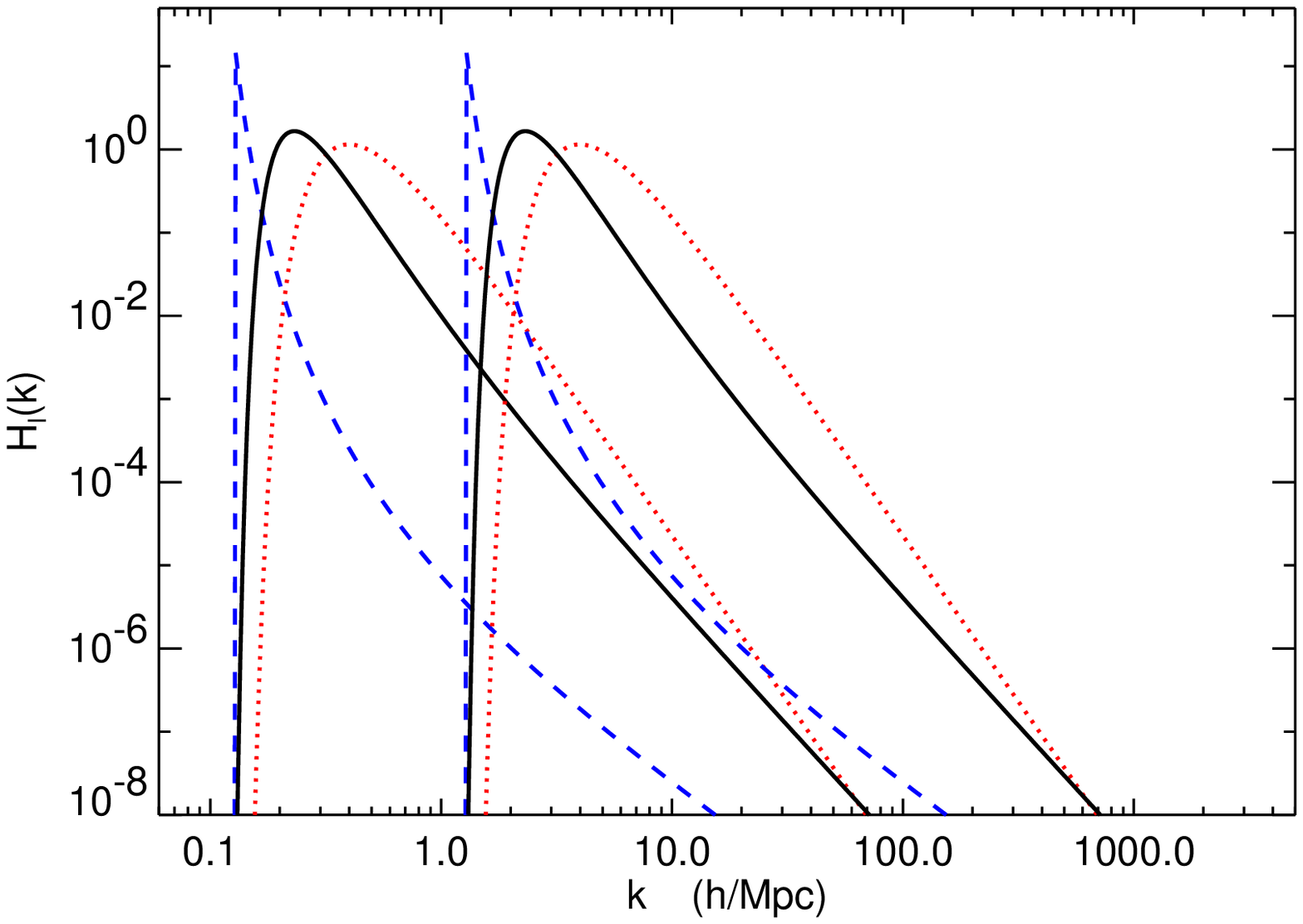,width=0.50\textwidth}
\epsfig{file=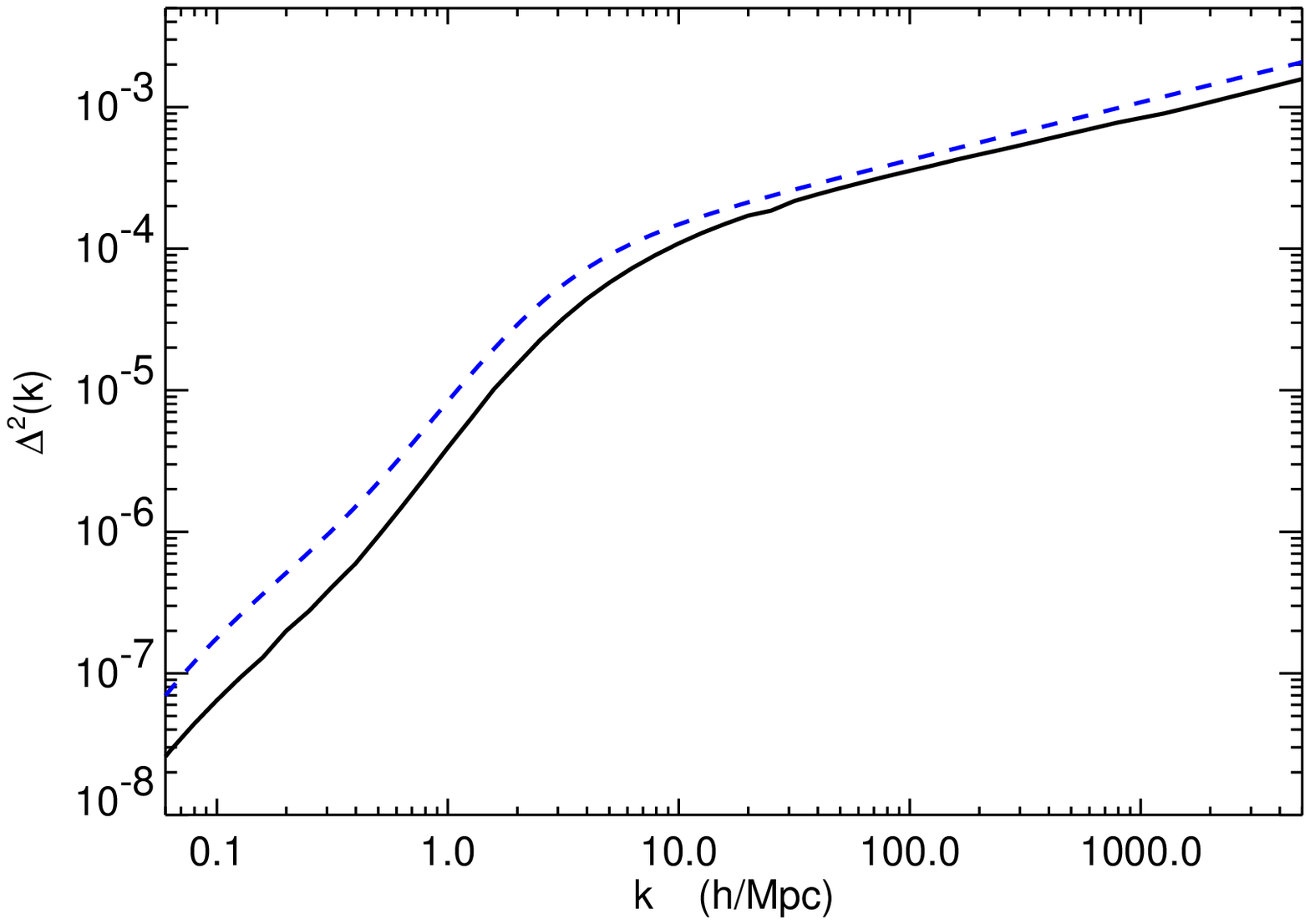,width=0.50\textwidth}}
\caption{Projection Kernels for the weak lensing kinetic-SZ correlation: 
The left panel shows the projection kernels $H_{\ell}\left(k\right)$ for the
the weak lensing auto correlation (dotted) $C^{\kappa\kappa}_{\ell}$, the  
kinetic-SZ auto correlation (dashed) $C^{\rm kSZ-kSZ}_{\ell}$, and the 
weak lensing-kinetic SZ squared cross-correlation 
(solid) $C^{\rm kSZ^2-\kappa}_{\ell}$. All kernels have been arbitrarily
 normalized to unit area.  The leftmost set of curves indicate 
the range of wave-numbers contributing to the angular power spectrum 
at $\ell=1000$; the rightmost set 
is for $\ell=10000$. The right panel shows dimensionless 3-d
power spectra of the radial momentum component (dashed) 
$\Delta^2_{p_{\hatn}p_{\hatn}}$ (eq.~\ref{MaFry}) and the triangular 
power spectrum (solid) $\Delta^2_{\cal T}$ (eq.~\ref{deltaT}). The triangle 
power spectrum is of the same order of 
magnitude as the power spectrum responsible for the kSZ effect, 
$\Delta^2_{{\cal T}} \sim \Delta^2_{p_{\hatn}p_{\hatn}}$}
\label{fig:hkern_KSZ}
\end{figure*}

The physical scales probed by these power spectra is 
illustrated in fig.~{\ref{fig:hkern_KSZ}, where we plot the projection 
kernels $H_{\ell}(k)$, for  $C^{\kappa\kappa}_{\ell}$, 
$C^{\rm kSZ-kSZ}_{\ell}$, and $C^{\rm kSZ^2-\kappa}_{\ell}$, for $\ell=1000$ 
and $10000$. The 3-d power spectrum of the radial momentum 
$\Delta^2_{p_{\hatn}p_{\hatn}}$ and the triangle power spectrum 
$\Delta^2_{{\cal T}}$ are plotted in the right panel at $z=0.5$. 
Note that the triangle power spectrum is of the same order of magnitude 
as the power spectrum responsible for the kSZ effect, 
$\Delta^2_{{\cal T}} \sim \Delta^2_{p_{\hatn}p_{\hatn}}$. 

The angular power spectrum is the projection of the power spectra on the 
right weighted by the kernels on the left (see eq.(~\ref{cross_k})), in this
figure.  The area under the $H_{\ell}(k)$ are normalized to unity. 
The sharp cutoff in the kernel for $C^{\rm kSZ-kSZ}_{\ell}$, occurs because
we have assumed the universe is reionized abruptly at 
$\eta(z_{\rm rei}=20)$ so that $W^{\rm kSZ}(\ell/k)$ drops to zero at small 
$k$.

We show the differential redshift contribution to each angular power 
spectrum as a function of redshift for $\ell=5000$ in fig.~\ref{fig:dlnz_KSZ},
where we plot 
$\frac{1}{\Delta^2_{XY}(\ell)}\frac{d\Delta^{2}_{XY}(\ell)}{dz}$,
for the auto and cross spectra. By definition, the area under each of 
these curves integrates to unity. While the kSZ auto power 
spectrum (dashed) has contributions from a high redshift tail, the bulk of 
the signal is skewed toward low z $\sim 0.5$, because of the growth of 
nonlinear structure. The overlap with the weak lensing (dotted) is thus
apparent, explaining the large correlation coefficient found in 
fig.~\ref{fig:d2l_kappa_KSZ2} (left).

\begin{figure}[t]
\centering
\centerline{\epsfig{file=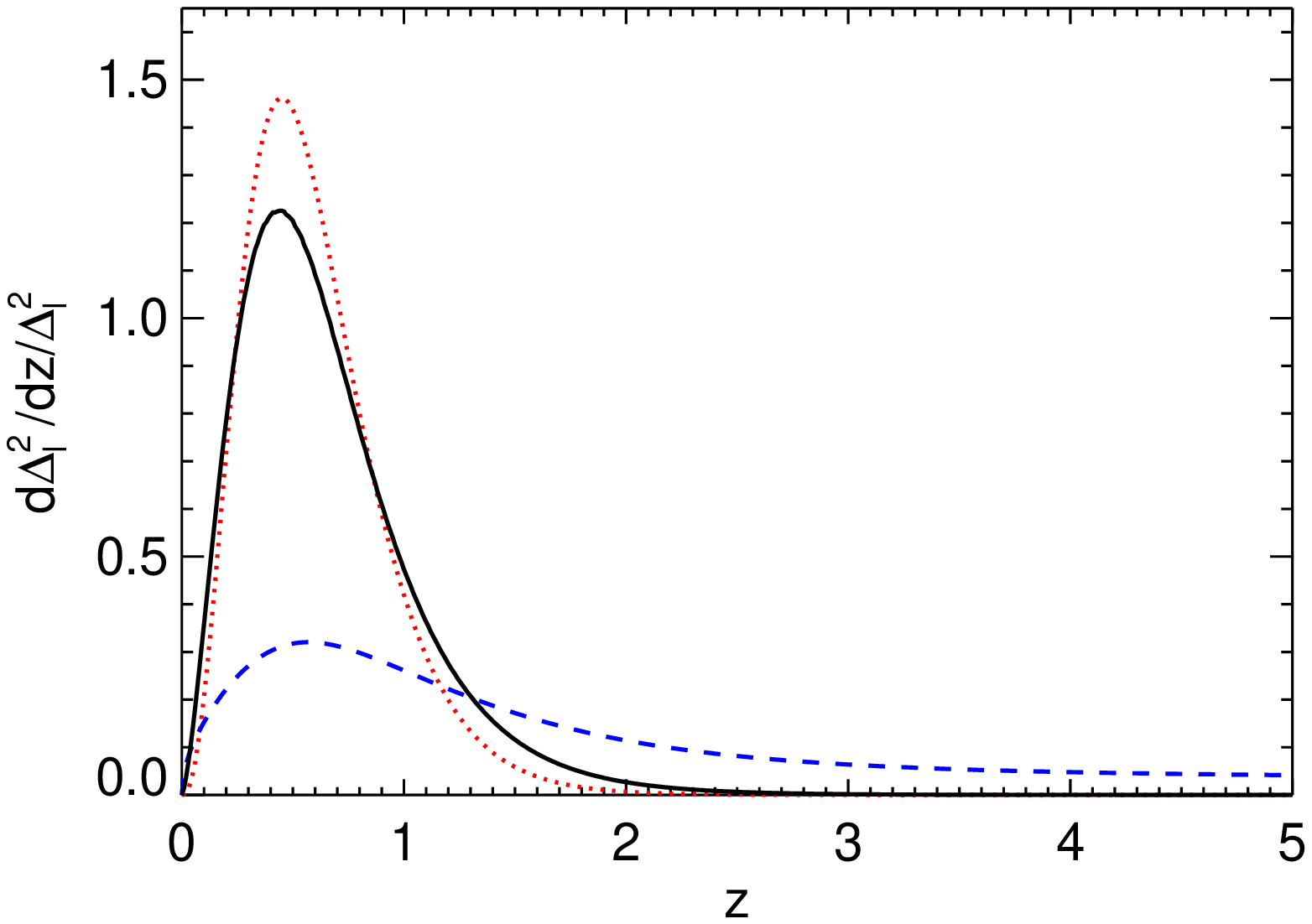,width=0.50\textwidth}}
\caption{Differential redshift contributions to angular power spectra: 
Differential redshift contributions to the weak lensing auto correlation 
$\frac{\ell^2}{2\pi}C^{\kappa\kappa}_{\ell}$ (dotted), the  kinetic-SZ 
auto correlation $\frac{\ell^2}{2\pi}C^{\Theta\Theta}_{\ell}$ (dashed), 
and the weak lensing-kinetic SZ squared cross-correlation 
$\frac{\ell^2}{2\pi}C^{\kappa\Theta^2}_{\ell}$ at $\ell=5000$.
All curves have been divided by the total power so that they 
integrate to unity.}\label{fig:dlnz_KSZ}
\end{figure}	

\section{Dependence on Power Spectrum Normalization}
\label{sec:sig8}
\begin{figure}[t]
\centering
\centerline{\epsfig{file=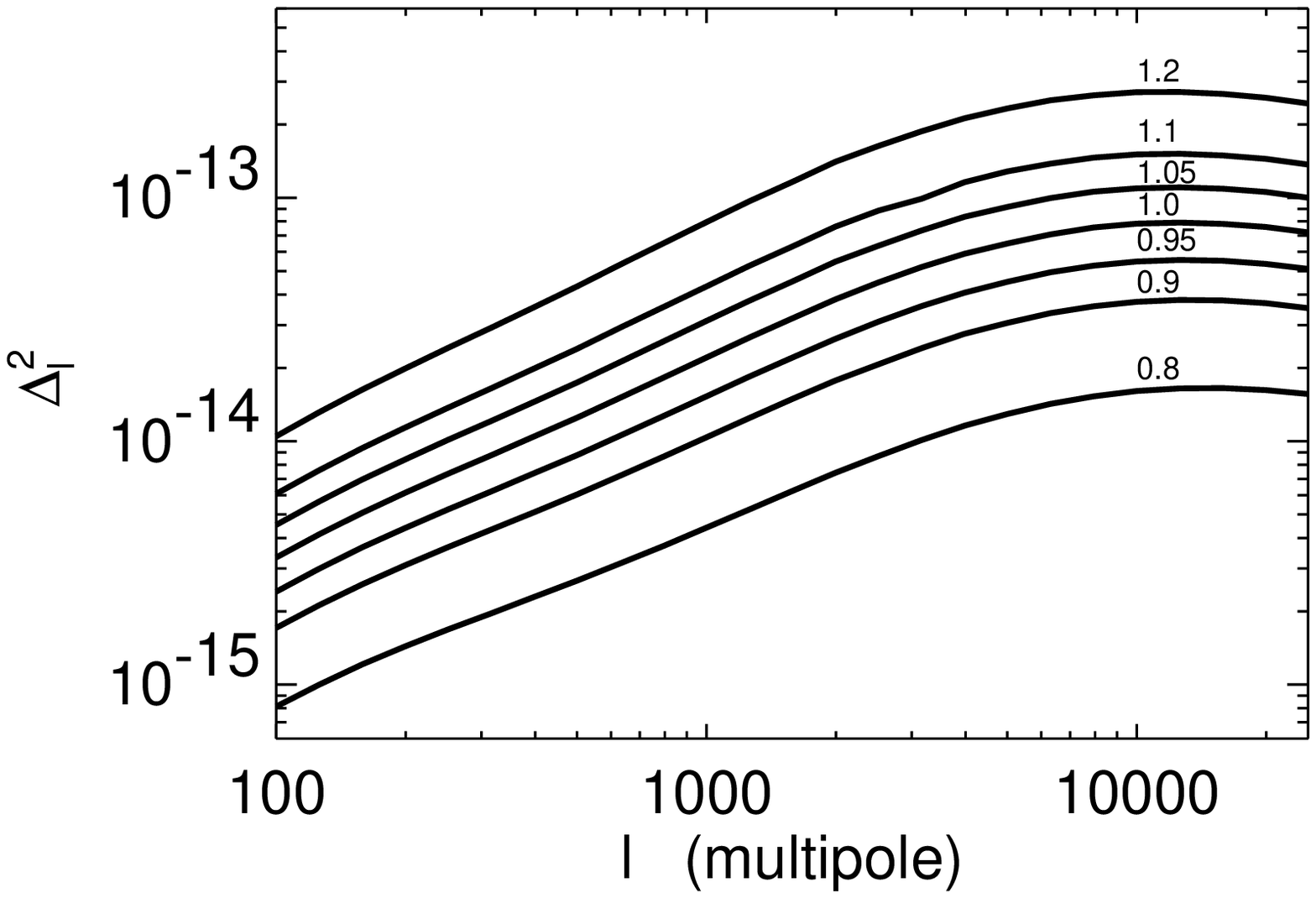,width=0.50\textwidth}}
\caption{Dependence of the kSZ$^2-\kappa$ cross correlation on $\sigma_8$. 
From top to bottom, the lines indicate $\sigma_8$=1.2, 1.1, 1.05, 1.0, 0.95, 0.9, and 0.8 as labeled in the figure.}\label{fig:sig8}
\end{figure}

It is well known that the thermal SZ power spectrum exhibits a strong
dependence on the power spectrum normalization parameter $\sigma_8$ 
\citep{Se01,ZhPen01,KoSe02}, approximately scaling as 
the seventh power $C_{\ell}\propto \sigma_8^7$. This strong dependence 
suggests that the thermal SZ power spectrum might be an effective probe of 
$\sigma_8$, because even if theoretical predictions are off by a factor of 2, 
this translates into less than 10\% systematic error in $\sigma_8$. 

We can determine the rough $\sigma_8$ dependence of the kSZ$^2$-weak 
lensing correlation from simple power counting if we recall that
$C^{\rm{kSZ^2-\kappa}_{\ell}} \propto v_{\rm{rms}}^2 B_{\rm{\delta\delta\delta}}$. The matter 
power spectrum scales as $P_{\delta\delta}^{} \propto \sigma_8^{2-3}$ --- two 
powers of $\sigma_8$ in the linear regime and three in the strongly nonlinear
stable clustering regime. The large scale modes that give rise to the bulk 
flows in  $v_{\rm rms}^2$ are linear, so we expect 
$v_{\rm rms}^2 \propto P^{\rm lin}_{\delta\delta} \propto \sigma_8^2$.
Finally, the bispectrum scales as $B_{\rm \delta\delta\delta}^{}\propto P_{\delta\delta}^2$, so putting
everything together we expect the scaling 
$C^{\rm kSZ^2-\kappa}_{\ell} \propto \sigma_8^{6-8}$. Figure~\ref{fig:sig8} 
illustrates the dependence of the cross power spectrum on $\sigma_8$. We find
that $C^{\rm kSZ^2-\kappa} \propto \sigma_8^7$ provides a good description of
this scaling. This is similar to the strong seventh power scaling of the 
thermal SZ and suggests that the kSZ$^2$-weak lensing correlation might also
serve as an effective probe of the power spectrum normalization and
its redshift evolution.

\section{Prospect for Measurement: Signal to Noise Analysis}
\label{sec:prospects}

In order to assess the detectability of the kSZ$^2$-weak lensing correlation
we consider several current and future CMB experiment 
configurations and WL lensing survey specifications

On the CMB side, we consider the future PLANCK mission and the soon to
be build ACT~\citep{ACT} telescope. Whereas the first one should offer a full sky
survey ($f_{\rm sky}\simeq 0.85$ if one takes into account galactic
contamination) the second one should conduct a smaller (100 square degree ($f_{\rm sky}\simeq
0.0025$)) but deeper survey. Both precise characteristics are
summarized in table 1, where $\theta_{\rm fwhm}$ denotes the beam full
width at half-maximum, $\sigma_{pix}$ denotes the instrumental
noise per $\theta_{\rm}^2$ area pixels and ``Area'' denotes the
observed sky area.

The subsequent expected errors are illustrated on 
fig.~\ref{fig:plot_clt}. Both have a frequency coverage appropriate
for a proper tSZ ``separation'' so that we neglect this signal from
now on (but the effects of residuals in \S~\ref{sec:prospects}). 
On the lensing side, we consider 3 various surveys: the
on-going Canada France Hawaii Telescope Legacy Survey (CFHTLS) 
\footnote{{\texttt http://www.cfht.hawaii.edu/Science/CFHTLS}}, the
future SNAP satellite \footnote{{\texttt http://snap.lbl.gov}} and
the future ground based Large Synoptic Survey Telescope (LSST) 
\footnote{{\texttt http://www.lssto.org}}. Their key characteristics
are summarized in table 2 and the expected
uncertainties are illustrated in fig.~\ref{fig:plot_clt}~:
 $z_0$ denotes the redshift parameter of the source distribution
of eq.~(\ref{p_z}), $n_{\rm gal}$ the mean density of galaxies, the
surveyed area and the \emph{single component} rms shear due to
intrinsic ellipticity is $\sigma_\gamma$.  

We define the optimal signal to noise per individual mode $\ell$ as 
\be
\biggr({S\over N}\biggl)_\ell^2 = {\bigl(C_\ell^{XY}\bigr)^2\over {\rm Cov}\bigr[\bigl(C_\ell^{XY}\bigr)^2\bigl]}\; .
\label{eq:sn_l}
\ee
To derive this formula, the assumption was made that all the relevant
fields were Gaussian so that we end up naturally with the $S/(S+N)$
Wiener optimal weighting. Doing so, we follow \eg \citet{Za00}. We then approximate the
covariance matrix by its diagonal, \ie neglecting the correlation induced by the cut sky,
\bea
\lefteqn{\bigl(f_{\rm sky}(2\ell+1)\bigr)^{-1}{\rm Cov}\bigr[\bigl(C_\ell^{XY}\bigr)^2\bigl]\quad = 
     }                       & \nonumber\\
& \bigl(C_\ell^{X} +  N_\ell^{X}\bigr)\bigl(C_\ell^{Y} + N_\ell^{Y}\bigr) + \bigl(C_\ell^{XY}\bigr)^2\ ,
\eea
where $N^{XX}_\ell$ denotes the evaluated instrumental noise contribution
to the measured angular power spectrum of $\ell$. For the weak lensing, it is
simply the shot noise due to the intrinsic ellipticitie of the sources, \ie 
\be
N_\ell^{WL} = \sigma_\gamma^2/n_{\rm gal}\ ,
\ee
whereas for the CMB, we include in the noise definition all the signal that can
not be separated out using multi-frequency information and the
conjunction of the noise and beam smearing give rise to \citep{Kn95} 
\be
N_{\ell\ \rm instr}^{\Theta} = \theta_{\rm fwhm}^2 \sigma_{\rm pix}^2/(4\pi)\ e^{\ell^2\theta_{\rm fwhm}^2/8\ln2}\ .
\ee

In the following subsections, we will consider two types of overlapping 
surveys. Either a small survey, like CFHTLS cross-correlated with
either ACT or SPT, in which case the overlap area is limited by the 
sky coverage of say ACT, which is 100 square degrees, or a very large survey 
which uses PLANCK in conjunction with LSST or PANSTAR. In
the latter case the total overlapping sky is limited by LSST, which 
is around 30,000 square degrees.

\begin{table}[b]
\begin{center}
\caption{Main characteristics of the CMB experiments discussed.}
\begin{math}
\begin{array}{c|c|c|c}
            & \theta_{\rm fwhm} & \sigma_{\rm pix}  & {\rm Area} \\
            &               & \left[\mu{\rm K}\right]       & \left[{\rm deg.}^2\right]  \\
  \hline\hline
  ACT / SPT\footnote{SPT has in fact a wider but shallower survey but
            the correlation would thus be limited by lensing survey.} &  1.7     &  2.                 & 100.   \\
  PLANCK    &  5.      &  2.2                & 34000. \\
  \hline
\end{array}
\end{math}
\end{center}
\label{tab:cmb_exp}
\end{table}

\begin{table}[b]
\begin{center}
\caption{Main characteristics of the weak-lensing surveys discussed.}
\begin{math}
\begin{array}{c|c|c|c|c}
  & z0 & n_{\rm gal}               & {\rm Area} & \sigma_{\gamma}\\
  &          &[\rm{gal.arcmin}^{-2}] & [\rm{deg.}^2] & \\
  \hline\hline
  CFHTLS  &  0.5     &  20.                 & 170.   & 0.31\\
  SNAP    &  0.9     &  100.                & 300.   & 0.23\\
  LSST    &  0.75    &  75.                 & 30000. & 0.16\\
\hline
\end{array}
\end{math}
\end{center}
\label{tab:lensing_exp}
\end{table}

\subsection{Signal to noise for the kSZ$^2$-WL correlation}

\begin{figure*}[t]
\centering
\centerline{\epsfig{file=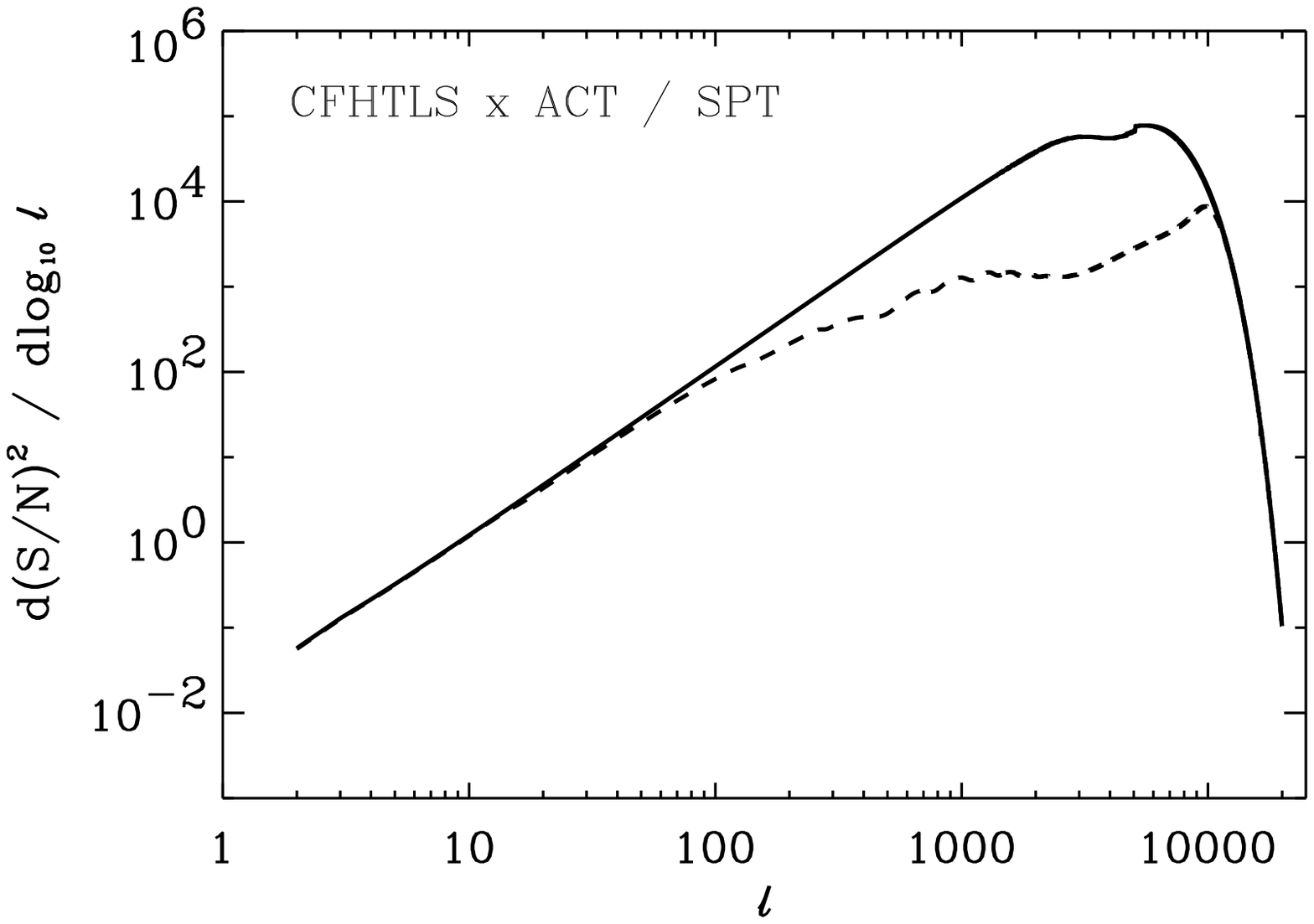,width=0.50\textwidth}\epsfig{file=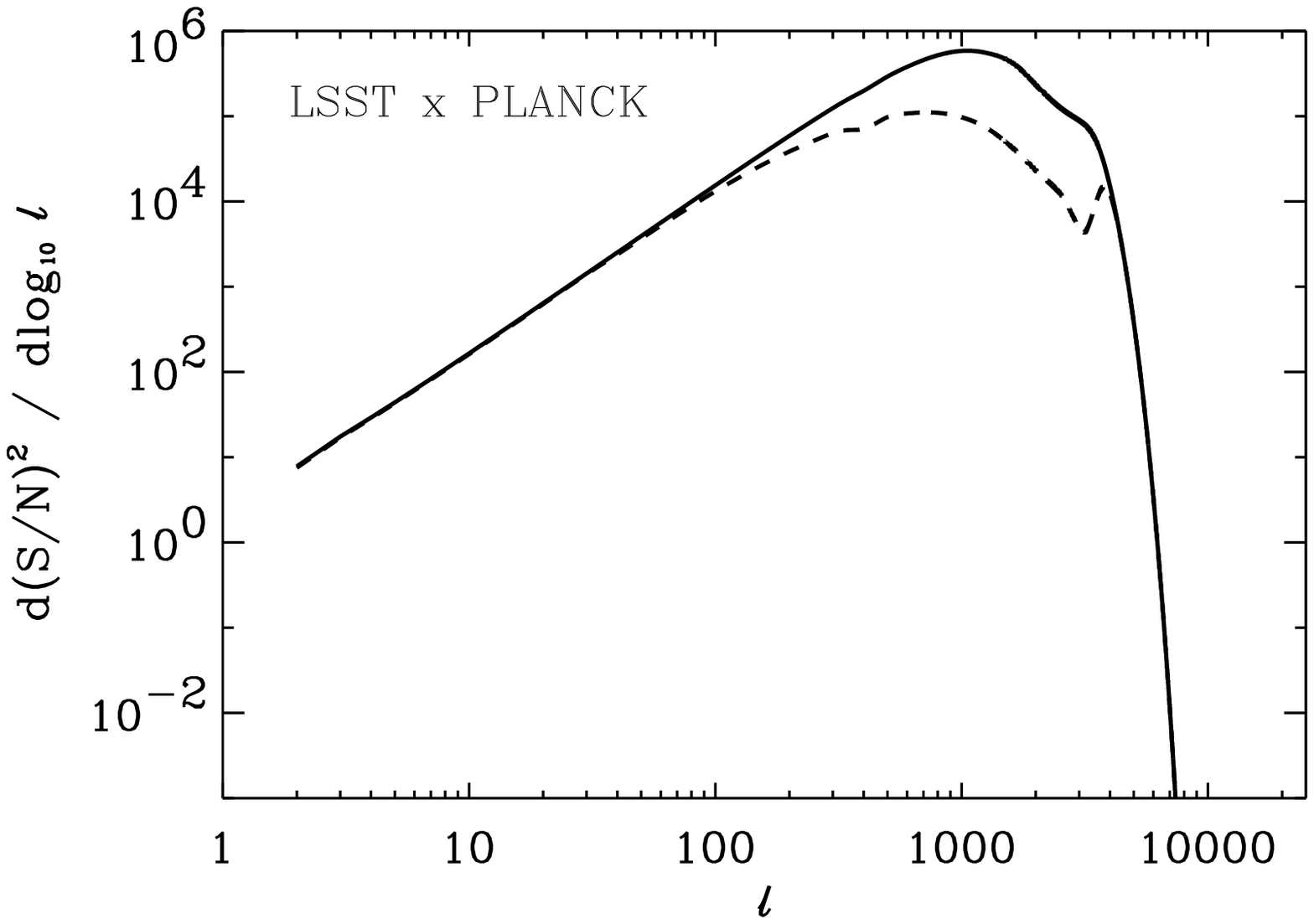,width=0.50\textwidth}}   
\caption{Signal to noise estimation for \emph{logarithmic bins} in $\ell$:
The left panel assumes an ideal full sky measurement free of any
instrumental noise. The middle panel illustrates the expectation for
a 100 square degree of a CFHTLS like survey and ACT. The right panel assumes a
LSST like 30 000 degree square degree and PLANCK. For each of those
plot, the dashed line corresponds to a worst case scenario where the
patchy reionisation contribution is 10 times higher (in temperature
square units) than the kSZ one.} \label{fig:ksz2wl_sn} 
\end{figure*}

For this correlation, the dominant ``noise'' contribution that need to
be considered at the relevant angular scales ($\sim 2^\prime$, \ie
$\ell\simeq 5000$, see fig.~\ref{fig:plot_clt}) are the lensed
primordial CMB, the potential patchy reionisation contribution and the
instrumental noise. We neglect any residual tSZ signal and the
subsequent correlation it would induced, we also neglect the
``spurious'' correlation due to lensing of the primordial 
CMB, although we discuss both of them in the next section. As is evident from
fig.~\ref{fig:plot_clt}, the amplitude of spurious correlation with 
lensing of the CMB should be be more than 2 orders of magnitude weaker. 

To evaluate both $C_\ell^{\rm kSZ^2\ kSZ^2}$ and $N_\ell^{kSZ^2}$ involves
one more approximation. We indeed have to evaluate the power spectrum
of quantities squared in real space. To do so exactly would involve
the computation of even higher order statistics, \eg it would be a 8
point function for $C_\ell^{\rm kSZ^2\ \rm kSZ^2}$. To avoid such
complications, we can reasonably consider those fields as being
weakly non-gaussian, so that we can neglect this way their connected
part, in particular the trispectrum of the squared field \citep{Za00}. With
this hypothesis, the evolution of the power spectrum of those
quantities translates in a convolution in harmonic space that can be
easily performed numerically~: 
\bea
C^{X^2}(\bfell) & \simeq & 2\int{d^2\bfell^\prime\over (2\pi)^2} C^{X}(\bfell^\prime)\ C^{X}(\bfell-\bfell^\prime)\label{eq:clconvol}\\
& \simeq & \int{\ell^\prime d\bfell^\prime\over \pi} C^{X}(\ell^\prime)\ C^{X}(|\bfell-\bfell^\prime|)\nonumber
\eea

We evaluate the convolution above numerically using the full power spectrum 
of primary plus secondary anisotropies. With this, we can compute all the required quantities in 
eq.~(\ref{eq:sn_l}). 

Before computing the theoretical prospects for the experimental
configurations we consider, it is crucial to realize that before
squaring the $\Theta$ field, an appropriate filtering is
necessary. Indeed, as visible \eg in eq.~(\ref{eq:clconvol}), this
squaring will introduce violent mode-coupling, so that any signal that
has a strong power outside the range where the kSZ signal dominates
will pollute the quadratic field. It would be possible using our
theoretical knowledge of the shape of the cross-correlation power
spectrum $C_\ell^{\kappa \Theta_{\rm kSZ}^2}\displaystyle$ , although a bit
cumbersome, to compute an optimal filter for our statistics \citep{Hu02}. 
Instead we take a reasonable short cut. Given the fact that 
both $C_\ell^\kappa$ and $C_\ell^{\kappa\Theta_{\rm kSZ}^2}$ are nearly 
flat in the range where the $kSZ$ dominates, we will consider indeed as 
the filter for our cross-correlation signal the one that maximizes 
the kSZ signal alone. This last one can be easily derived if we assume that the
projected kSZ signal is weakly non-Gaussian. It is indeed given by the
usual Wiener filter \citep{BoGi99,TeEf96}  
\be
f_\ell = {C_\ell^{\rm kSZ} \over C_\ell^{\rm Noise} + C_\ell^{\rm kSZ}}
\ee
where $C_\ell^{\rm Noise}$ includes the instrumental noise but also all
the other astrophysical component different from the kSZ
itself. Note that this filter is similar to the one derived in a
more general manner by \citet{Hu02} and employed by \citet{Co01a}.  

The results are illustrated in fig.~\ref{fig:ksz2wl_sn} where we
plot the contribution to the square of signal to noise per $\log \ell$. We
consider two situations: i/ the dashed line corresponds to a situation
where the patchy reionisation is present and strong, \ie we consider
the higher amplitude model of \citep{Sa03}, where the patchy
reionisation power spectrum amplitude is about 10 times higher than
the kSZ one in the same angular range (see fig.~\ref{fig:plot_clt})
ii/ the solid line corresponds to a scenario where the patchy
reionisation component is completely negligible. 

Obviously, this signal should be easily detectable using a CFHTLS-ACT
configuration. The overall signal to noise ratio reaches 225 if the
patchy reionization is negligible, 47 otherwise. The signal to noise
per individual $\ell$ reaches 1.8. For the PLANCK-LSST configuration,
the signal reaches 978 without no patchy reionization and 473 otherwise. 
However, as we discuss in the next section, this kSZ signal will probably be
heavily contaminated by thermal SZ residuals. 

\subsection{Spurious Correlations}

The primary limitation to detecting the kSZ$^2$-weak lensing correlation 
will certainly come from spurious sources of correlation. Possible sources 
of anisotropy other than kSZ that the weak lensing may correlate with are 
tSZ residuals from imperfect frequency cleaning, weak lensing of the 
primordial CMB, and point sources. We discuss each in turn. 

The dominant spurious correlation will probably be that due to the residual 
tSZ after frequency cleaning. \citet{CoHuTeg00} estimated that for 
PLANCK, the power spectrum of the residual for $600<\ell<2500$ should be 
around 0.14 times the initial tSZ power spectrum, \ie around 5 times the kSZ
for our model. Since the correlation coefficient around WL and tSZ is
expected to be around 0.6~~\citep{Se01}, the correlated component between 
tSZ residuals and WL will probably be more significant than the correlation
with kSZ.  For ACT/SPT telescope, the frequency coverage will allow for 
frequency cleaning as good as if not better than PLANCK, while the
higher resolution provides improved sensitivity at the scales
where the tSZ dominates,$\ell >2000$ (see fig.~\ref{fig:plot_clt}), so 
frequency cleaning should be drastically improved at those scales. Further 
consideration of this issue is required, but we defer this to a future work. 

Another source of spurious correlation should be the correlation induced
between the lensed primordial CMB and the WL. Here again, this effect
might be important for PLANCK as visible in 
fig.~\ref{fig:plot_clt} but should be negligible for ACT/SPT 
since in the range of interest, the lensed contribution of the CMB is 
around 2 orders of magnitude weaker than the kSZ one. In addition, 
any attempt to cross correlate CMB lensing with  cosmic shear \citep{Hu02} 
will need to consider the kSZ$^2$-weak lensing signal as a source of 
spurious correlation, since this signal will begin to dominate for 
$\ell \sim 3000$ near the damping tail. 

Finally, there is the possibility of spurious correlation with point sources.
At the frequencies of interest, this is primarily dust emission 
from sub-mm galaxies. This population of objects will probably 
correlate with the weak lensing signal but the strength of that correlation
is rather uncertain.  

Finally, we note that the kSZ$^2$-weak lensing correlation has a 
specific signature that may alleviate some of the problems with 
spurious correlation and could provide a mechanism to isolate the kSZ 
signal. Specifically, the kSZ effect does not correlate with 
weak lensing at the 2 point level, whereas tSZ residuals or dusty sub-mm 
galaxies will. It may be possible to exploit this fact, and remove the part of 
the anisotropy that correlates with weak lensing at the two point level. This
would  effectively  ``lensing clean'' the temperature map so that the 
residuals would be dominated by the kSZ. Realistic simulations are 
certainly required to precisely evaluate the potential of this technique,  
but this is outside the scope of this paper.

\section{Conclusion}
\label{sec:discuss}

We have evaluated the prospects for upcoming experiments to measure
the correlation between arc-minute scale secondary anisotropies of the CMB 
and cosmic shear,  after tSZ has been removed by frequency cleaning. 
In particular, we evaluated the signal to noise ratio for the kSZ$^2$-weak lensing
correlation.  

The two point correlation of kSZ with and density tracer 
is negligible because of the isotropy of the velocity field; however, 
we found that a strong 3-point 
correlation exists. A collapsed three point statistic was introduced to 
measure the three point signal, which is the angular cross power spectrum 
between weak lensing and the square of the filtered temperature. 
While residual thermal SZ will limit the detectability of this correlation 
for PLANCK-LSST, it should be easily detectable by cross correlation 
ACT/SPT with the CFHTLS/SNAP at $\ell \sim 5000$ where the signal to noise 
ratio per $\ell$ reaches 1.8. In principle, a full analysis of the cross 
bispectrum $B^{\kappa\Theta\Theta}_{\ell_{1}\ell_{2}\ell_{3}}$ could yield a 
higher signal to noise --- a calculation we defer to a future paper. 

The kSZ$^2$-weak lensing correlation probes three point correlations between 
the underlying dark matter and the the \emph{momentum} of the ionized 
baryons in dense regions, providing the only known probe into this physics. 
We presumed that these higher order correlations, or more precisely the 
hybrid bispectrum $B_{\delta p_{\hatn}p_{\hatn}}$,  arose from three point 
density modulations of large scale coherent motions. The situation is surely
more complicated and this approximation must be compared to hydrodynamic 
numerical simulations.  Nevertheless, measurement of kSZ$^2$-weak lensing
correlation will provide valuable insights into the physics of energy 
injection, the bias and ionization fraction of baryons in the densest 
environments, and information on the nonlinear mode coupling between dark 
matter and baryons.  We also considered the dependence of correlation signal
on the power spectrum normalization parameter $\sigma_8$ and determined the 
approximate scaling $C_{\ell} \propto \sigma_8^7$ similar to the scaling of 
the tSZ power spectrum.  

Despite the interest in detection of the kSZ$^2$-weak lensing correlation, 
a null detection could prove equally interesting. The contribution of patchy 
reionization to arc-minute scale anisotropy in the CMB is a contentious 
issue, with different reionization models and histories predicting signal 
amplitudes differing by several orders of magnitude 
\citep{Ag96,GrHu98,KnSc98,Va01,Sa03}. The lack of a correlation in the 
presence of significant arc-minute scale anisotropy would indicate that this 
anisotropy was generated at high redshift, rather than in 
the low redshift universe probed by weak gravitational lensing. This last 
point is especially important when one considers the similarity in shape 
between the power spectrum of patchy reionization and that of kinetic SZ 
(see fig.~\ref{fig:plot_clt}). 

In addition, we note that many of the calculations in this paper 
can be directly applied to correlations with galaxies rather than weak 
lensing as the tracer
of density field, given a suitable model for the bias of galaxies in the 
nonlinear regime, as provided by, \eg the recently popular  halo model 
\citep{CoSh02,Sh03}. The weak 
lensing window function $W^{\kappa}(\eta)$, would simply be replaced 
with a suitable window function  for the galaxies. As the noise in 
the correlation measurement is dominated 
by the CMB, the signal to noise per square degree should not change 
significantly. However, because large area galaxy surveys already 
exist, the correlations we discussed might be measured with galaxies first, 
as was the case recently for the ISW effect 
\citep{boughn03,Nolta03,FoGa03,fosalba03,Sc03,As03}. 
The kSZ$^2$-galaxy correlation  would probe three point correlations 
between the density of galaxies and the momentum of ionized baryons in 
clusters, providing valuable information about galaxy formation.

With such a high signal to noise ratio, 
it is conceivable that radial information 
might also be extracted from the cross-correlation. If photometric redshifts 
of background source galaxies were available, one could attempt to 
deproject the cross-correlation tomographically. Similarly, if galaxies were 
correlated with kSZ$^2$ rather than weak lensing, galaxy photo-z's could be
use to extract this radial information. This has been discussed in the 
context of galaxy-tSZ cross correlation by \citet{ZhPen01}. Studying the 
cross-correlation as a function of redshift would yield 
information on the redshift evolution of the dynamical state of baryons in 
dense environments.  

In conclusion, our study predicts that coming arc-minute scale measurements
of the secondary anisotropies of the CMB should will correlate strongly
with local tracers of the density field, like weak gravitational lensing or 
galaxies from large scale structure surveys. The total signal to noise ratio
for the kSZ$^2$-weak lensing correlation is greater than 220 for 
ACT/SPT correlated with CFHTLS/SNAP, which will be easily detectable. We 
believe that this signal along with the other cross-correlations with cosmic 
shear studied by \cite{Se01} and \citet{Hu02}, provide a significant 
incentive for upcoming CMB experiments and lensing surveys to image 
the same regions of sky. 

\acknowledgments
We thank Eiichiro Komatsu for providing the thermal SZ power spectrum
and numerous helpful discussions. We acknowledge useful discussions with  
Lloyd Knox on patchy reionization, and we are grateful to Ravi Sheth 
for communicating unpublished results. We are grateful to Masahiro
Takada for pinpointing some corrupted units in the first version of
this paper.

\end{document}